\documentclass[11pt]{article}
\newcommand{\ours}{\mbox{SweepMark}\xspace}

\usepackage[preprint]{acl}

\usepackage{times}
\usepackage{latexsym}

\usepackage[T1]{fontenc}

\usepackage[utf8]{inputenc}

\usepackage{microtype}

\usepackage{inconsolata}

\usepackage{graphicx}
\usepackage{subfigure}

\usepackage{algorithm}
\usepackage{algorithmic}

\usepackage{amsfonts}
\usepackage{multirow}
\usepackage{amsmath}
\usepackage{xspace}
\usepackage{booktabs}

\title{SweepMark: Query-Free and High-Fidelity Attacks on Robust Audio Watermarking}

\author{
 \textbf{Weikang Ding\textsuperscript{1}},
 \textbf{Binhao Ma\textsuperscript{1}},
 \textbf{Hanqing Guo\textsuperscript{2}},
 \textbf{Rui Duan\textsuperscript{1}},
\\
\\
 \textsuperscript{1}University of Missouri-Kansas City,
 \textsuperscript{2}Indiana University Bloomington
\\
}

\begin{document}
\maketitle
\begin{abstract}
Audio watermarking embeds imperceptible signals into speech to trace misuse in text-to-speech (TTS) synthesis. Existing removal attacks may retain watermark-related artifacts or degrade speech quality when suppressing them. We propose \ours, a query-free black-box attack that targets these artifacts without access to target watermarking models or paired clean and watermarked samples. Our key insight is that watermark embedding introduces spectral patterns that are not fully aligned with natural speech, including persistent structures that conventional speech enhancement may retain. Diverse Artifact Learning extracts complementary stationary and non-stationary patterns, using customized spectral artifacts added to public speech to learn persistent structures without target watermark examples. Adaptive Artifact Scaling then combines and amplifies these patterns for subtraction while constraining distortion. Across two speech datasets and four watermarking methods, \ours achieves average attack success rates of 92\% and 96\%, with higher perceptual quality scores overall than the evaluated adaptive attacks. Automatic speech recognition evaluation indicates that linguistic content is largely preserved. Further experiments with three TTS systems show high attack success against two synthesis-resistant watermarking methods in speech generated after fine-tuning on attacked recordings.
\end{abstract}

\section{Introduction}

With the rapid advancement of voice synthesis technologies, text-to-speech (TTS) tools~\cite{shen2018natural,kong2020hifi} can be misused to impersonate a target speaker, potentially causing financial losses and reputational damage. Audio watermarking provides a proactive protection mechanism by embedding an imperceptible signal into speech for subsequent recovery by a dedicated decoder~\cite{liu2023detecting,roman2024proactive}. To defend against TTS misuse, methods such as Timbre~\cite{liu2023detecting} and AudioMarkNet~\cite{zong2025audiomarknet} seek to preserve watermark detectability in speech synthesized by models trained on watermarked recordings, despite changes in linguistic content and temporal structure.

Existing watermark removal attacks~\cite{liu2024audiomarkbench,wen2025sok} broadly fall into two categories based on their access requirements. Query-based attacks~\cite{lopez2024speech,li2025harmonicattack} repeatedly query the watermarking model to obtain paired clean and watermarked samples for training a dedicated removal model. These approaches incur substantial query costs and often generalize poorly to unseen watermarking schemes. Query-free attacks instead apply generative waveform reconstruction without interacting with the target watermarking model~\cite{yao2026audio,zhong2026vulnerability}. These operations may leave watermark-related components intact, while increasing their intensity to remove the watermark may also degrade speech quality.

These limitations raise a natural question: can we exploit the structure of TTS-resistant watermarks to disrupt them while preserving speech quality? We observe that methods such as Timbre and AudioMarkNet embed watermark information across speech frames to reduce dependence on temporal alignment, producing persistent spectral patterns~\cite{liu2023detecting,zong2025audiomarknet}. Existing query-free attacks do not explicitly target these patterns. This observation motivates us to design a method that disrupts watermark-related spectral structure while preserving linguistic content and speaker characteristics~\cite{ding2026learning}.



Guided by this observation, we propose SweepMark, a query-free audio watermark removal attack that suppresses both stationary and non-stationary watermark-related spectral artifacts. SweepMark adopts a two-stage framework, beginning with Diverse Artifact Learning to capture artifacts with different temporal characteristics. Specifically, a stationary branch learns to suppress persistent spectral patterns using a training dataset constructed by adding customized, temporally consistent spectral artifacts to clean speech. This design enables the branch to learn such patterns without requiring samples from the target watermarking models. A complementary non-stationary branch learns to suppress artifacts that vary with time and speech content. Together, the two branches support watermark removal across diverse spectral structures.

To remove watermarks while maintaining high perceptual fidelity, Adaptive Artifact Scaling adjusts the strength and combination of the extracted stationary and non-stationary patterns for each input. Directly subtracting these patterns may attenuate watermark evidence without causing decoding failure, as their watermark-related components are initially weak. We therefore adaptively scale and combine the two patterns to strengthen watermark suppression while limiting changes to the underlying speech. The resulting combined pattern is subtracted from the watermarked waveform, enabling \ours to accommodate sample-specific removal while preserving perceptual quality.

We make the following contributions:

\begin{itemize}
\item We develop a stationary artifact learning method supported by a customized dataset of clean speech augmented with temporally consistent spectral artifacts. This design enables the model to learn persistent watermark-related patterns without requiring samples from the target watermarking models.
\item We propose \ours, a query-free attack framework for high-fidelity audio watermark removal. Diverse Artifact Learning captures complementary stationary and non-stationary patterns, while Adaptive Artifact Scaling adjusts their strength and combination for each input to disrupt watermark recovery.
\end{itemize}

\section{Background and Related Work}

\subsection{Audio Watermarking}

Neural audio watermarking commonly adopts an encoder--decoder architecture to embed and recover watermark information~\cite{liu2024audiomarkbench,wen2025sok}. Its design balances perceptual fidelity and robustness: the embedded signal should be imperceptible and remain recoverable after audio processing. Specifically, the watermark encoder $E$ embeds a binary message $m$ into an audio waveform $x$, producing $x_w=E(x,m)$. This process can be represented as adding a watermark waveform $w$, i.e., $x_w=x+w$. In the time-frequency domain, let $\mathrm{STFT}(\cdot)$ denote the short-time Fourier transform (STFT). We define the magnitude spectrogram as $S=|\mathrm{STFT}(x)|$. A watermark pattern $\Phi_w$ of the same dimension gives the watermarked spectrogram $S_w=S+\Phi_w$. This additive representation describes watermark-induced modifications without requiring the encoder to operate explicitly in the STFT domain.

\subsection{Watermarking for Speech Synthesis}

TTS converts text into speech through acoustic modeling and waveform generation~\cite{shen2018natural,ren2020fastspeech,kong2020hifi}, or a jointly trained framework~\cite{kim2021conditional}. Speaker-adaptive systems such as YourTTS~\cite{casanova2022yourtts} further synthesize a target speaker's voice using reference recordings or speaker-specific fine-tuning.

Watermarking can be applied before or after synthesis. Post-generation methods, such as WavMark~\cite{chen2023wavmark} and AudioSeal~\cite{roman2024proactive}, embed watermarks into existing waveforms for detection or attribution. Robustness to audio processing does not necessarily ensure survival through subsequent speech synthesis~\cite{wen2025sok}. In contrast, Timbre~\cite{liu2023detecting} and AudioMarkNet~\cite{zong2025audiomarknet} seek to preserve watermark detectability when watermarked recordings are used for voice cloning. New input texts change linguistic content, duration, and temporal structure, requiring watermark evidence to persist despite these changes. This requirement motivates examining the temporal characteristics of watermark-related spectral patterns.

\subsection{Audio Watermark Removal Attacks}

Watermark removal attacks aim to disrupt detection or decoding while preserving audio utility. We organize existing methods by their attack knowledge.

\paragraph{White-box attacks.}
These attacks exploit target model parameters or gradients to optimize watermark-evading modifications~\cite{liu2024audiomarkbench,ding2026learning}. Their applicability is limited when model internals are unavailable.

\paragraph{Query-based black-box attacks.}
Detector-query attacks use prediction feedback to search for adversarial modifications, employing techniques such as HopSkipJumpAttack~\cite{chen2020hopskipjumpattack} and Square Attack~\cite{andriushchenko2020square}. Encoder-query attacks collect clean--watermarked pairs to train removal models~\cite{lopez2024speech,li2025harmonicattack}. Both require interaction with the target system, potentially incurring substantial query costs and service restrictions. Training on scheme-specific pairs can additionally limit transfer to unseen watermarking methods.

\paragraph{Query-free attacks.}
These attacks operate without querying the target encoder or detector. Signal-processing approaches apply transformations such as filtering and compression~\cite{wen2025sok,ozer2025comprehensive}. Speech enhancement suppresses components treated as noise using conventional or learned models~\cite{chen2006new,pascual2017segan,richter2023speech}, and prior studies have examined its effectiveness for watermark removal~\cite{lopez2024speech,o2025deep}. Generative approaches instead reconstruct the waveform through learned audio models~\cite{ozer2026self,zhong2026vulnerability}; DiffErase~\cite{yao2026audio}, e.g., combines diffusion with a vocoder. These operations may retain watermark-related components or alter speech characteristics as processing strength increases. In contrast, \ours learns complementary stationary and non-stationary artifact patterns and adaptively combines them to suppress watermark evidence while preserving perceptual quality.

\section{Methodology}
\label{sec:method}

\begin{figure*}[t]
    \centering
    \includegraphics[width=\linewidth]{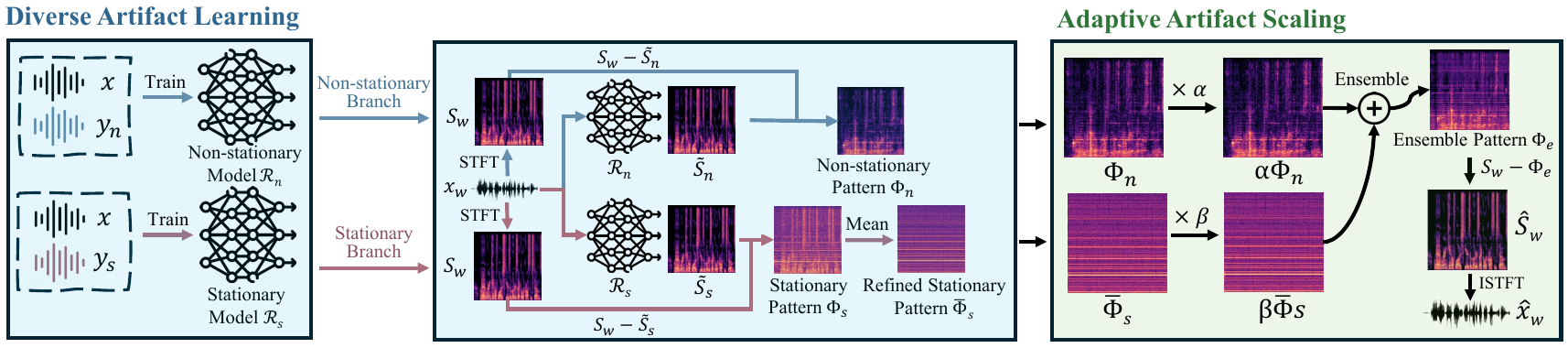}
    \caption{Overview of the two-stage attack framework of \ours. Diverse Artifact Learning extracts non-stationary and stationary patterns; Adaptive Artifact Scaling adaptively scales and combines them to construct an ensemble pattern.}
    \label{fig:framework}
\end{figure*}

\subsection{Threat Model}
The attacker removes the watermark from
watermarked audio while preserving speaker characteristics and intelligibility. The attacker can access publicly available speech corpora, but has no access to clean-reference audio from the target speaker, clean-watermarked pairs from the target watermarking model, or the watermark encoder, decoder, and detector, and cannot query watermark information. The attacker also has no knowledge of the specific watermarking method deployed by the defender.

\subsection{Motivation and Overview}
To make audio watermarks both imperceptible and robust, watermarking models usually embed watermark signals into audio regions that satisfy human psychoacoustic constraints. However, the embedding process still introduces additional patterns into the spectrogram. These patterns are important for watermark decoding, but they are not necessarily consistent with the underlying speech structure. Applying sufficiently strong perturbations or directly reconstructing the waveform may disrupt the original speech structure and introduce unnecessary artifacts. Therefore, our goal is to suppress watermark-related components while preserving the integrity of the underlying speech. Speech enhancement offers a way to suppress such components using models trained on publicly available speech, but it still faces two challenges.

\noindent\textbf{Limited Artifact Diversity.}
Existing speech enhancement methods are mainly designed to remove audible environmental noises. However, watermark signals can appear as non-speech artifacts with diverse structural characteristics, including stationary and non-stationary patterns in the spectrogram. Since such artifact patterns are not sufficiently covered by existing methods, off-the-shelf speech enhancement models may fail to capture them effectively across different watermarking schemes.

\noindent\textbf{Incomplete Artifact Suppression.}
Speech enhancement models are primarily optimized to remove perceptually salient distortions that degrade speech quality. They tend to suppress only the most prominent non-speech artifacts, while weaker artifacts are often preserved. Effective removal requires stronger suppression of these components while retaining the underlying speech structure.

We address these two challenges with an operator that targets non-speech artifacts directly rather than the unknown watermark. The design maps one component to each challenge, as Figure~\ref{fig:framework} shows.

\noindent\textbf{Stage 1: Diverse Artifact Learning.}
We employ two models with complementary design objectives to capture diverse non-speech artifacts. Specifically, the framework learns complementary artifact patterns corresponding to non-stationary and stationary artifact structures, improving the coverage of artifact characteristics introduced by different watermarking schemes.

\noindent\textbf{Stage 2: Adaptive Artifact Scaling.}
Since enhancement models mainly suppress perceptually prominent artifacts, the extracted artifact patterns may not completely cover weak abnormal components. Therefore, we adaptively scale and combine the extracted artifact patterns under a quality-preserving constraint, enabling broader artifact suppression while maintaining speech quality.





\subsection{Diverse Artifact Learning}

Diverse Artifact Learning employs two complementary models to capture non-speech artifacts with different structural characteristics, which are categorized into two types: non-stationary artifacts and stationary artifacts. Non-stationary artifacts vary over time and appear as localized distortions in the time-frequency domain, whereas stationary artifacts remain relatively stable along the temporal axis and appear as frequency-persistent structures.

Accordingly, we employ two enhancement models with complementary design objectives. The non-stationary enhancement model $\mathcal{R}_n$ extracts a non-stationary artifact pattern $\Phi_n$, while the stationary enhancement model $\mathcal{R}_s$, trained on a customized stationary watermark-like noise dataset, extracts an initial stationary artifact pattern $\Phi_s$. Since stationary artifacts may appear temporally discontinuous, we further apply a temporal mean operation to each frequency bin of $\Phi_s$ to obtain the refined stationary artifact pattern $\bar{\Phi}_s$.

\noindent\textbf{Non-stationary pattern generation branch.}
We train a conditional diffusion model $\mathcal{R}_n$ to output a non-stationary pattern $\Phi_n$. The pattern captures non-stationary artifacts that are inconsistent with the distributional structures of clean audio samples. First, we collect a training dataset that contains the clean and non-stationary pairs. Given the clean audio waveform $x$ and non-stationary noise waveform $n_n$, we generate the corresponding noisy audio waveform $y_n$ by injecting $n_n$ into $x$: $y_n=x+n_n$.
To train the model $\mathcal{R}_n$, we sample $x_t$ at time step $t$ as follows: $x_t=\mu(x_0,y_n,t)+\sigma(t)z$,
where $\mu$ and $\sigma$ are defined following~\cite{sarkka2019applied, richter2023speech}, and $z \sim \mathcal{N}_{\mathbb{C}}(0, I)$. The training objective is:
\begin{equation}
\arg\min_{\theta}
\mathbb{E}
\left[
\left\lVert
s_{\theta}(x_t,y_n,t)
+
\frac{\mathbf{z}}{\sigma(t)}
\right\rVert_2^2
\right].
\label{eq:training_objective_nonstationary_conditional_diffusion}
\end{equation}
The non-stationary pattern $\Phi_n$ is expressed as:
\begin{equation}
\Phi_n=S_{w}-\widetilde{S}_n, \quad \widetilde{S}_n=|\mathrm{STFT}(\mathcal{R}_n(x_w))|.
\label{eq:nonstationary_pattern}
\end{equation}

\noindent\textbf{Stationary pattern generation branch}. Compared with the non-stationary pattern branch, we make two modifications: (1) we construct stationary noise samples and use this clean-noisy paired dataset for model training; and (2) we apply an additional mean operation to the generated stationary pattern. We first construct a customized noise dataset specifically for watermark removal, where imperceptible stationary noise is injected into the clean audio samples. Then, we train a conditional diffusion model $\mathcal{R}_s$ to extract a stationary pattern $\Phi_s$, which captures stationary components from the spectrogram. Moreover, we apply a mean operation along the time axis to repair the discontinuous stationary structure for each frequency bin and generate a refined stationary pattern $\bar{\Phi}_s$.

It is non-trivial to construct a clean-noisy paired dataset. In real-world scenarios, watermarks are imperceptible to human hearing, and real stationary watermark patterns are unavailable for model training. Therefore, inspired by existing noisy dataset construction strategies~\cite{botinhao2016investigating,richter2024ears}, we construct stationary noise samples in the time-frequency domain and scale them based on signal-to-noise ratio (SNR) values. First, we transform the clean audio waveform $x$ into the time-frequency domain $S$. Second, we randomly select several frequency bins $\mathcal{F}$ as valid noise injection targets. For each selected frequency bin $f$, we compute its average spectrogram $\bar{S}$ across all time frames $L$ and use this value to construct the stationary noise spectrogram $S_n$ while preserving the original phase information. After applying this operation to all selected frequency bins, we transform the constructed noise spectrogram $S_n$ back to the time domain using ISTFT. Third, to control the noise level and generate the stationary noise waveform $n_s$, we use a scale coefficient $r$ to construct the noise waveform $n$ with a random SNR value $\rho$. Finally, we add this stationary noise waveform $n_s$ to the clean audio waveform $x$ and then generate a stationary-noise audio waveform $y_s=x+n_s$. The Algorithm~\ref{alg:stationary_noise} shows the process. In this paper, the number of selected frequency bins ranges from 10 to 50, and the SNR ranges from 25 to 35 dB. We train $\mathcal{R}_s$ using the same training objective as $\mathcal{R}_n$, with $y_n$ replaced by $y_s$:
\begin{equation}
\arg\min_{\theta}
\mathbb{E}
\left[
\left\lVert
s_{\theta}(x_t,y_s,t)
+
\frac{\mathbf{z}}{\sigma(t)}
\right\rVert_2^2
\right].
\label{eq:stationary_model_training}
\end{equation}
The stationary pattern $\Phi_s$ is computed as:
\begin{equation}
\Phi_s=S_{w}-\widetilde{S}_s, \quad \widetilde{S}_s=|\mathrm{STFT}(\mathcal{R}_s(x_w))|.
\label{eq:stationary_pattern}
\end{equation}

After obtaining $\Phi_s$, we apply an additional mean operation to generate the refined stationary pattern $\bar{\Phi}_s$. From our observation of $\Phi_s$, we find that some stationary noise components are not temporally continuous. Therefore, to better capture the statistical properties of the stationary structure, we use a mean operation on $\Phi_s$ and obtain a refined noisy pattern $\bar{\Phi}_s$, which is expressed as:
\begin{equation}
\begin{array}{c}
\displaystyle
\bar{\Phi}_s(f,l)
=
\frac{1}{L}
\sum_{\tau=1}^{L}
\Phi_s(f,\tau), \\
\displaystyle
f=1,\ldots,F,\; l=1,\ldots,L .
\end{array}
\label{eq:stationary_pattern_mean}
\end{equation}

\subsection{Adaptive Artifact Scaling}

For artifacts located in imperceptible regions, the two models often suppress them rather than completely remove them. Therefore, the extracted patterns may contain weak watermark-related artifacts. In this section, we aim to amplify the patterns generated by the two models under a quality-preserving constraint.

Given a non-stationary pattern $\Phi_n$ and a refined stationary pattern $\bar{\Phi}_s$, we combine them to construct an ensemble pattern $\Phi_e$. The ensemble pattern is defined as:
\begin{equation}
\Phi_e=\alpha \Phi_n + \beta\bar{\Phi}_s,
\label{eq:ensemble_pattern}
\end{equation}
where $\alpha$ and $\beta$ are the scaling parameters for $\Phi_n$ and $\bar{\Phi}_s$, respectively.

Since the energy of $\Phi_n$ and $\bar{\Phi}_s$ is dynamic across different audio samples, the $\alpha$ and $\beta$ should not be fixed values. To dynamically determine these scaling parameters, we propose a two-step optimization strategy. In the first step, we scale the refined stationary pattern $\bar{\Phi}_s$ to a fixed SNR value, such as 35 dB, which determines $\beta$. In the second step, we keep $\beta \bar{\Phi}_s$ unchanged and only adjust the scaling parameter $\alpha$ for the non-stationary pattern $\Phi_n$. We combine the scaled non-stationary pattern $\alpha \Phi_n$ with the fixed stationary pattern $\beta \bar{\Phi}_s$, and search for a suitable $\alpha$ to make the ensemble pattern $\alpha \Phi_n + \beta \bar{\Phi}_s$ reach another target SNR value, e.g., 20 dB, which determines $\alpha$. In this way, the stationary component is first amplified to a controlled level, and the non-stationary component is then adaptively adjusted based on the already fixed stationary component. This strategy ensures the ensemble pattern is adaptively amplified according to the energy distribution of each audio sample.

Finally, we subtract the ensemble pattern $\Phi_e$ from the watermarked spectrogram $S_w$ to obtain the attacked spectrogram $\hat{S}_w=S_w-\Phi_e$, and then apply ISTFT to $\hat{S}_w$ to obtain the attacked audio waveform $\hat{x}_w$.

\section{Evaluation}

\begin{table*}[htbp]
\centering
{\small
\setlength{\tabcolsep}{0.6mm}{
\resizebox{\linewidth}{!}{%
\begin{tabular}{llcccccccccc}
\toprule
\multirow{2}{*}{\textbf{\shortstack{Attack\\Type}}}
& \multirow{2}{*}{\textbf{Attack Method}}
& \multicolumn{5}{c}{\textbf{Flickr dataset}}
& \multicolumn{5}{c}{\textbf{WSJ0 dataset}} \\
\cmidrule(lr){3-7}
\cmidrule(lr){8-12}
& 
& \textbf{AudioSeal}
& \textbf{WavMark}
& \textbf{Timbre} 
& \textbf{AudioMarkNet}
& \textbf{Avg.}
& \textbf{AudioSeal}
& \textbf{WavMark}
& \textbf{Timbre} 
& \textbf{AudioMarkNet}
& \textbf{Avg.} \\
\midrule

\multirow{3}{*}{\textbf{SP}}
& NoiseReduce   & 0.82 & 0.94 & 1.00 & 1.00 & 0.94 & 0.33 & 0.74 & 1.00 & 0.50 & 0.64 \\
& Wiener Filter & 0.00 & 0.00 & 0.02 & 0.00 & 0.01 & 0.00 & 0.00 & 0.00 & 0.00 & 0.00 \\
& MMSE\_STSA    & 0.06 & 0.00 & 0.00 & 0.27 & 0.08 & 0.00 & 0.00 & 0.00 & 0.18 & 0.05 \\
\midrule

\multirow{3}{*}{\textbf{SE}}
& MP-SENet      & 0.14 & 0.02 & 0.01 & 0.96 & 0.28 & 0.06 & 0.09 & 0.03 & 0.99 & 0.29 \\
& MetricGAN+    & 0.12 & 0.06 & 0.00 & 0.27 & 0.11 & 0.02 & 0.00 & 0.00 & 0.10 & 0.03 \\
& StoRM         & 0.18 & 0.15 & 0.04 & 0.79 & 0.29 & 0.15 & 0.32 & 0.01 & 1.00 & 0.37 \\
\midrule

\multirow{6}{*}{\textbf{Adaptive}}
& BigVGAN       & 1.00 & 1.00 & 0.82 & 0.19 & 0.75 & 1.00 & 1.00 & 0.88 & 0.12 & 0.75 \\
& Square        & 0.19 & 0.87 & 0.12 & 0.00 & 0.30 & 0.14 & 0.96 & 0.36 & 0.06 & 0.38 \\
& HSJA          & 0.98 & 0.97 & 1.00 & 0.98 & 0.98 & 0.70 & 0.90 & 1.00 & 0.96 & 0.89 \\
& DiffErase-M & 1.00 & 1.00 & 0.95 & 0.19 & 0.79 & 1.00 & 1.00 & 0.94 & 0.14 & 0.77 \\
& DiffErase-L & 1.00 & 1.00 & 0.96 & 0.38 & 0.84 & 1.00 & 1.00 & 0.95 & 0.30 & 0.81 \\
& \textbf{Ours} & \textbf{0.84} & \textbf{0.91} & \textbf{0.94} & \textbf{1.00} & \textbf{0.92}
                & \textbf{0.85} & \textbf{0.99} & \textbf{1.00} & \textbf{1.00} & \textbf{0.96} \\
\bottomrule

\end{tabular}}
}
}
\vspace{-2mm}
\caption{Watermark Attack Success Rate (WASR) against different audio watermarking methods. The attack types include three categories: Signal Processing (SP) attacks, Speech Enhancement (SE) attacks, and Adaptive attacks.}
\label{tab:attack_results}
\end{table*}

\begin{table*}[t]
\centering
{\small
\setlength{\tabcolsep}{0.8mm}
\resizebox{\linewidth}{!}{%
\begin{tabular}{lcccccccccccc}
\toprule
\multirow{2}{*}{\textbf{\shortstack{Attack\\Method}}}
& \multicolumn{3}{c}{\textbf{AudioSeal}}
& \multicolumn{3}{c}{\textbf{WavMark}}
& \multicolumn{3}{c}{\textbf{Timbre}}
& \multicolumn{3}{c}{\textbf{AudioMarkNet}} \\

\cmidrule(lr){2-4}
\cmidrule(lr){5-7}
\cmidrule(lr){8-10}
\cmidrule(lr){11-13}

& \textbf{NISQA}
& \textbf{DNSMOS}
& \textbf{UTMOS}
& \textbf{NISQA}
& \textbf{DNSMOS}
& \textbf{UTMOS}
& \textbf{NISQA}
& \textbf{DNSMOS}
& \textbf{UTMOS}
& \textbf{NISQA}
& \textbf{DNSMOS}
& \textbf{UTMOS} \\
\midrule

w/o attack
& 3.87 & 3.19 & 4.17
& 4.05 & \underline{3.23} & 4.10
& 3.64 & 3.18 & 4.03
& 3.36 & 2.77 & 4.01 \\
\midrule

BigVGAN
& 3.78 & 3.06 & 3.49
& 3.91 & 3.07 & 3.44
& 3.58 & 3.08 & 3.25
& 3.78 & 3.01 & 3.66 \\

Square
& 1.71 & 2.35 & 2.32
& 1.74 & 2.38 & 2.13
& 1.68 & 2.36 & 2.18
& 1.85 & 2.36 & 2.57 \\

HSJA
& 1.20 & 1.81 & 1.55
& 2.41 & 2.70 & 3.42
& 1.16 & 1.85 & 1.38
& 1.09 & 1.54 & 1.42 \\

DiffErase-M
& 3.56 & 3.02 & 3.28
& 3.72 & 3.02 & 3.24
& 3.43 & 3.04 & 3.13
& 3.82 & 3.02 & 3.44 \\

DiffErase-L
& 3.75 & 3.15 & 3.38
& 3.96 & 3.18 & 3.37
& 3.64 & 3.17 & 3.29
& 3.94 & 3.18 & 3.37 \\
\midrule

\textbf{Ours}
& \textbf{\underline{4.22}} & \textbf{\underline{3.25}} & \textbf{\underline{4.25}}
& \textbf{\underline{4.29}} & \textbf{3.22} & \textbf{\underline{4.18}}
& \textbf{\underline{4.28}} & \textbf{\underline{3.23}} & \textbf{\underline{4.10}}
& \textbf{\underline{4.59}} & \textbf{\underline{3.31}} & \textbf{\underline{4.22}} \\

\bottomrule

\end{tabular}}
}
\vspace{-2mm}
\caption{Audio quality evaluation under adaptive attacks on the WSJ0 dataset. w/o attack is watermarked audio without attack.}
\label{tab:audio_quality}
\end{table*}

\subsection{Experimental Setup}

\noindent\textbf{Dataset.} We mainly use three datasets in our experiments: one for training and two for inference. The training and inference datasets have no overlap. We use VoiceBank-DEMAND~\cite{botinhao2016investigating} as the training dataset. For inference, we use the Flickr 8k Audio Caption Corpus~\cite{harwath2015deep} and the Wall Street Journal (WSJ0)~\cite{garofolo1993csr}.

\noindent\textbf{Audio Watermarking Methods.} We select four state-of-the-art watermarking methods: AudioSeal~\cite{roman2024proactive}, WavMark~\cite{chen2023wavmark}, Timbre~\cite{liu2023detecting}, and AudioMarkNet~\cite{zong2025audiomarknet}.

\noindent\textbf{Evaluation Metrics.} We use the \textbf{Watermark Attack Success Rate (WASR)} to measure \emph{how successfully the attackers perform the watermark attack}. We use the default rules for each watermark detector. For audio quality evaluation, we use three MOS-based metrics: NISQA~\cite{mittag2021nisqa}, the overall quality score (OVRL) of DNSMOS P.835~\cite{reddy2021dnsmos,reddy2022dnsmos}, and UTMOS~\cite{saeki2022utmos}.

\noindent\textbf{Automatic Speech Recognition.} We use three ASR models: SpeechBrain~\cite{ravanelli2021speechbrain} with CRDNN trained on LibriSpeech, Wav2Vec2.0~\cite{baevski2020wav2vec} Base fine-tuned on 960hr LibriSpeech, and Whisper~\cite{radford2023robust} Base.

\noindent\textbf{Text-to-Speech.} For datasets, we follow the default settings of the original papers: LJSpeech dataset in Timbre~\cite{ljspeech17} and former US president Obama dataset in AudioMarkNet~\cite{zong2025audiomarknet}. For TTS models, we use Tacotron 2~\cite{shen2018natural}, FastSpeech 2~\cite{ren2020fastspeech}, HiFi-GAN~\cite{kong2020hifi} as the vocoder, and YourTTS~\cite{casanova2022yourtts}.

\noindent\textbf{Audio Watermark Attack Methods.} We compare our method with eleven watermark removal attacks from three categories. The first category is \emph{signal processing attacks}, including NoiseReduce~\cite{sainburg2020finding}, Wiener Filter, MMSE\_STSA ~\cite{ephraim1984speech}. The second category is \emph{speech enhancement attacks}, including MP-SENet~\cite{lu2023mp}, MetricGAN+~\cite{fu2021metricgan+}, StoRM~\cite{lemercier2023storm}. The third category is \emph{adaptive attacks}, including BigVGAN~\cite{lee2022bigvgan}, the HSJA-Signal attack~\cite{chen2020hopskipjumpattack} and Square attack~\cite{andriushchenko2020square} implemented by AudioMarkBench~\cite{liu2024audiomarkbench}, and DiffErase-Mel and DiffErase-Latent~\cite{yao2026audio}. To eliminate the potential influence of data discrepancy, we train these models on the same dataset used by our method, i.e., VoiceBank-DEMAND. In our experiments, since SGMSE+ can generate audio with high perceptual quality~\cite{richter2023speech}, we use the model architecture of SGMSE+ by default. Supplementary~\ref{sec:additional_experiment_setput_settings} shows additional experimental setup details.

\subsection{Attack Performance}

Table~\ref{tab:attack_results} shows the attack results. For signal processing (SP) attacks, only NoiseReduce achieves a high WASR, but it sacrifices much audio quality, with an average NISQA score of 3.34 and an average UTMOS score of 3.39.

For speech enhancement (SE) attacks, most attacks fail to achieve effective watermark removal, with most WASR values below 0.2. One exception is that MP-SENet and StoRM achieve relatively high WASR against AudioMarkNet, especially on the WSJ0 dataset. This is because AudioMarkNet embeds watermarks below 1000 Hz, which can be more easily identified and suppressed by speech enhancement attacks.

For adaptive attacks, most methods achieve high WASR, and our attack demonstrates strong performance across all four watermarking methods, with average WASRs of 0.92 and 0.96 on two datasets, respectively. Additionally, among the five existing attacks, only HSJA successfully attacks AudioMarkNet. This indicates that models trained with BigVGAN and DiffErase are less effective at handling watermark artifacts in low-frequency regions. Compared with the other attacks, our WASR against AudioSeal is slightly lower, although the average WASR still exceeds 0.9. The goal of our attack is to achieve successful watermark removal while preserving audio quality. Existing attacks can successfully remove watermarks in some cases, but they destroy more speech structure.

\begin{figure}[t]
    \centering
    \includegraphics[width=\columnwidth]{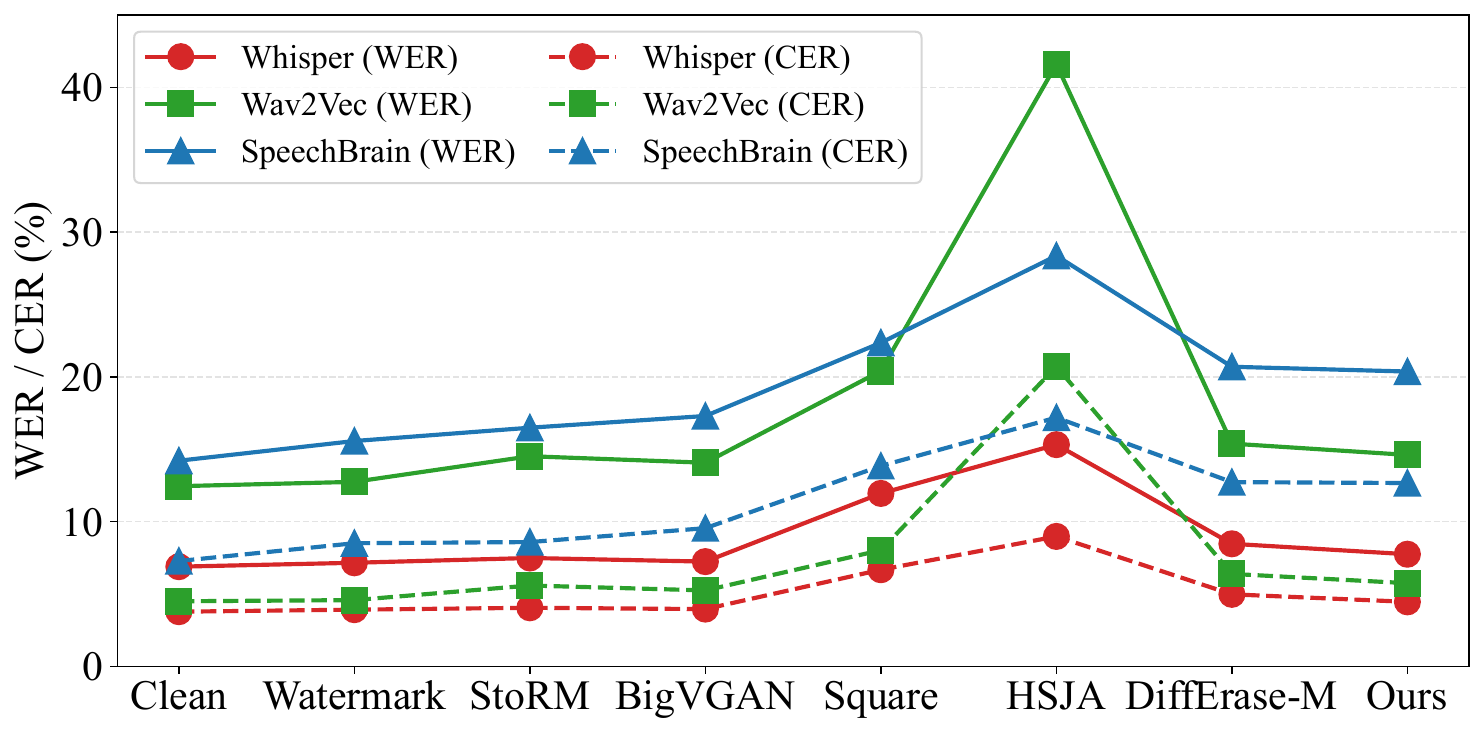}
    \caption{WER (\%) and CER (\%) evaluated on three ASR models. Timbre is the watermarking method.}
    \label{fig:automatic_speech_recognition}
\end{figure}

\subsection{Audio Quality}

Our goal is to suppress perceptual artifacts while preserving speech content. Reference-based metrics such as SNR measure the energy difference between the reference audio and the attacked audio, but they do not always accurately reflect human perception~\cite{sawata2022diffiner,hirano2026diffusion}. Therefore, we evaluate audio quality under adaptive attacks using three metrics: NISQA, DNSMOS, and UTMOS. Supplementary~\ref{sec:additional_experiment_on_audio_quality} presents the complete results and additional analysis of reference-based metrics.

The results are shown in Table~\ref{tab:audio_quality}. Our attack preserves audio quality better than the existing adaptive attacks. We observe that HSJA achieves high WASR but has poor perceptual quality. BigVGAN and DiffErase, which directly reconstruct audio using trained models, improve audio quality compared with Square attack and HSJA. However, they still fail to preserve the original audio quality, and their overall quality is even lower than that of the watermarked audio without attack. Since the training and inference datasets do not overlap, models trained on one dataset may not generalize well to another dataset. In contrast, our attack is designed to balance attack effectiveness and audio quality. Rather than directly reconstructing the audio using a model, our method optimizes and scales the extracted patterns while preserving the original audio phase. In this way, our method suppresses watermark-related regions while better maintaining the speech structure and perceptual quality.

\begin{table}[t]
    \centering
    \setlength{\tabcolsep}{2pt}
    \resizebox{\linewidth}{!}{%
    \begin{tabular}{clcccccc}
        \toprule
        \multirow{2}{*}{TTS model}
        & \multirow{2}{*}{Attack method}
        & \multicolumn{3}{c}{Timbre}
        & \multicolumn{3}{c}{AudioMarkNet} \\
        \cmidrule(lr){3-5} \cmidrule(lr){6-8}
        & & WASR & UTMOS & WER & WASR & UTMOS & WER \\
        \midrule

        \multirow{7}{*}{\shortstack{FastSpeech2\\+HiFi-GAN}}
        & Clean     & 1.00 & 4.10 & 6.97 & 0.93 & 2.30 & 21.13   \\
        & Watermark & 0.00 & 4.13 & 5.96 & 0.00 & 2.29 & 22.44 \\
        & Square & 0.92 & 3.45 & 6.50 & 0.02 & 2.03 & 31.28 \\
        & HSJA & 0.95 & 3.45 & 6.56 & 0.76 & 1.24 & 54.01 \\
        & DiffErase-M & 1.00 & 3.99 & 6.85 & 0.09 & 2.44 & 20.77 \\
        & DiffErase-L & 1.00 & 3.83 & 6.67 & 0.57 & 2.23 & 19.21 \\
        & \textbf{Ours}      & \textbf{1.00} & \textbf{4.12} & \textbf{6.20} & \textbf{0.99} & \textbf{2.75} & \textbf{19.87} \\
        \midrule

        \multirow{7}{*}{\shortstack{Tacotron2\\+HiFi-GAN}}
        & Clean     & 1.00 & 4.28 & 5.24 & 0.91 & 2.01 & 20.38   \\
        & Watermark & 0.00 & 3.86 & 6.08 & 0.01 & 2.06 & 20.73 \\
        & Square & 1.00 & 3.01 & 6.85 & 0.01 & 1.80 & 23.54 \\
        & HSJA & 1.00 & 3.19 & 6.20 & 0.39 & 1.27 & 83.51 \\
        & DiffErase-M & 1.00 & 4.00 & 6.38 & 0.02 & 2.08 & 24.26 \\
        & DiffErase-L & 1.00 & 3.63 & 5.48 & 0.53 & 1.69 & 21.81 \\
        & \textbf{Ours}      & \textbf{1.00} & \textbf{4.11} & \textbf{5.42} & \textbf{1.00} & \textbf{2.53} & \textbf{19.90} \\
        \midrule

        \multirow{7}{*}{YourTTS}
        & Clean     & 1.00 & 3.60 & 18.06 & 0.96 & 2.69 & 43.39 \\
        & Watermark & 0.02 & 3.48 & 18.12 & 0.00 & 2.69 & 43.15 \\
        & Square & 0.95 & 2.81 & 32.42 & 0.01 & 2.44 & 47.62 \\
        & HSJA & 0.88 & 2.86 & 30.39 & 0.17 & 1.28 & 110.00 \\
        & DiffErase-M & 1.00 & 3.10 & 25.45 & 0.05 & 2.28 & 44.70 \\
        & DiffErase-L & 1.00 & 3.07 & 24.43 & 0.12 & 2.63 & 43.27 \\
        & \textbf{Ours}   & \textbf{1.00} & \textbf{3.67} & \textbf{17.58} & \textbf{0.99} & \textbf{3.09} & \textbf{38.08} \\

        \bottomrule
    \end{tabular}%
    }
    \vspace{-2mm}
    \caption{TTS synthesis performance evaluation. Whisper outputs the WER.}
    \label{tab:tts_parts_results}
\end{table}

\subsection{Automatic Speech Recognition Evaluation}

Figure~\ref{fig:automatic_speech_recognition} shows the average WER and CER of three ASR models evaluated on two datasets under Timbre, which provides an assessment of whether the linguistic content is preserved after watermark removal attacks. From the results, we can observe that the impact of our attack on the ASR models remains within a reasonable range~\cite{shah2025speech}, with average increases of approximately 3.06 percentage points in WER and 2.44 percentage points in CER compared with clean audio on Timbre. Compared with other attack methods, our attack achieves lower WER and CER than the two query-based attacks, Square and HSJA, and performs comparably to the other attacks. Our attack can effectively preserve linguistic content while maintaining high watermark removal performance. Supplementary~\ref{sec:additional_experiment_on_automatic_speech_recognition_evaluation} shows the complete results.

\subsection{Text-to-Speech Analysis}

Table~\ref{tab:tts_parts_results} evaluates how attacked speech affects subsequent TTS-synthesized speech in terms of attack effectiveness, audio quality, and linguistic content. For each attack, we fine-tune the TTS model on the attacked watermarked samples and generate 100 utterances for evaluation. Clean and Watermark denote fine-tuning on clean and unmodified watermarked samples, respectively.

Our attack achieves the highest WASR, which indicates that our method remains effective after TTS synthesis. Moreover, compared with other attacks, our method achieves higher UTMOS and relatively lower WER, demonstrating better preservation of audio quality and linguistic content. Additional experimental details and results are provided in Supplementary~\ref{sec:addtional_experiment_on_text_to_speech_analysis}.
\begin{figure}[t]
    \centering
    \includegraphics[width=0.98\columnwidth]{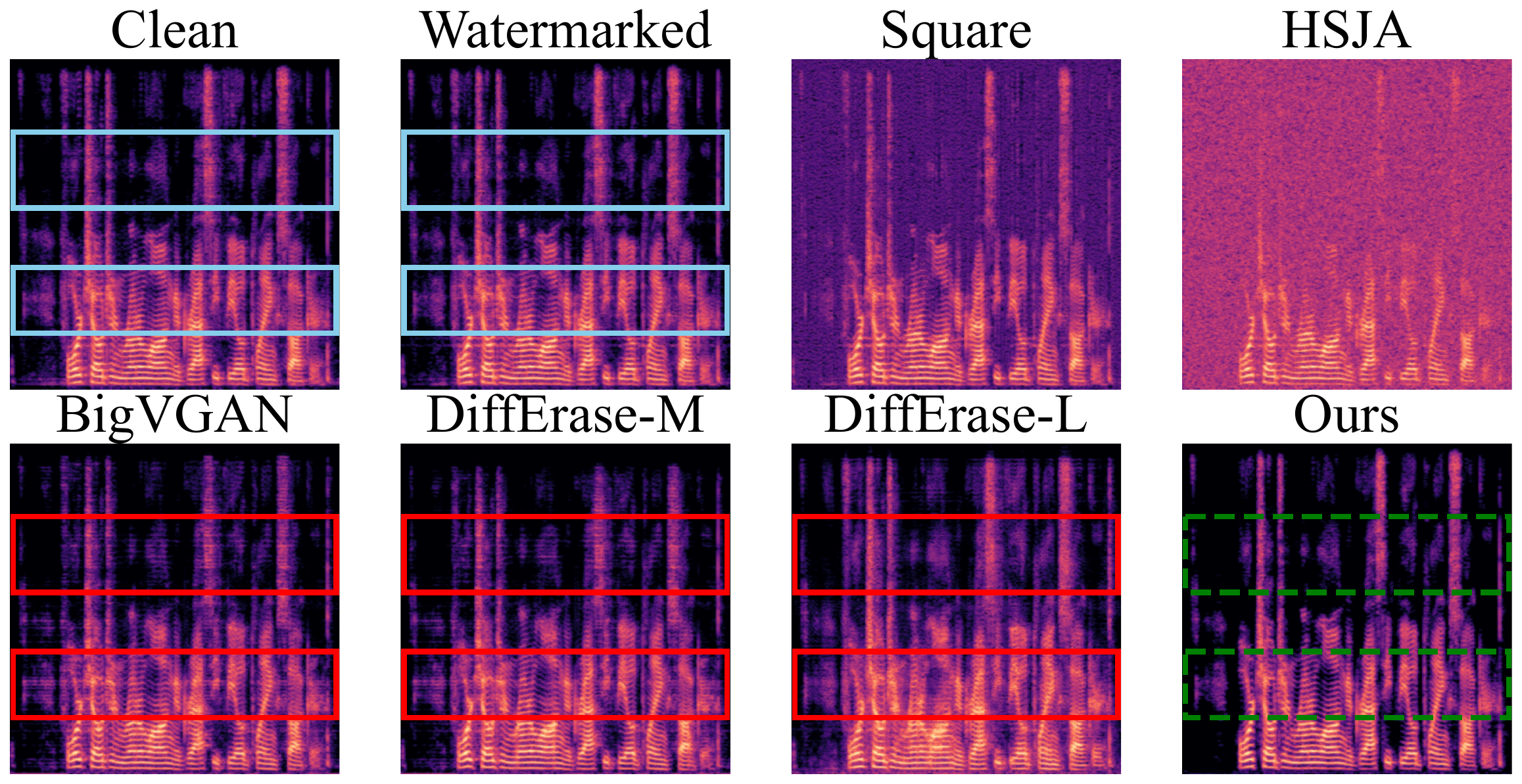}
    \caption{Attacked audio visualization. The red solid boxes in BigVGAN and DiffErase highlight regions of noticeable distortion. Timbre is used as the watermarking method.}
    \label{fig:audio_visualization_audioseal}
\end{figure}

\begin{figure}[t]
    \centering
    \includegraphics[width=0.98\columnwidth]{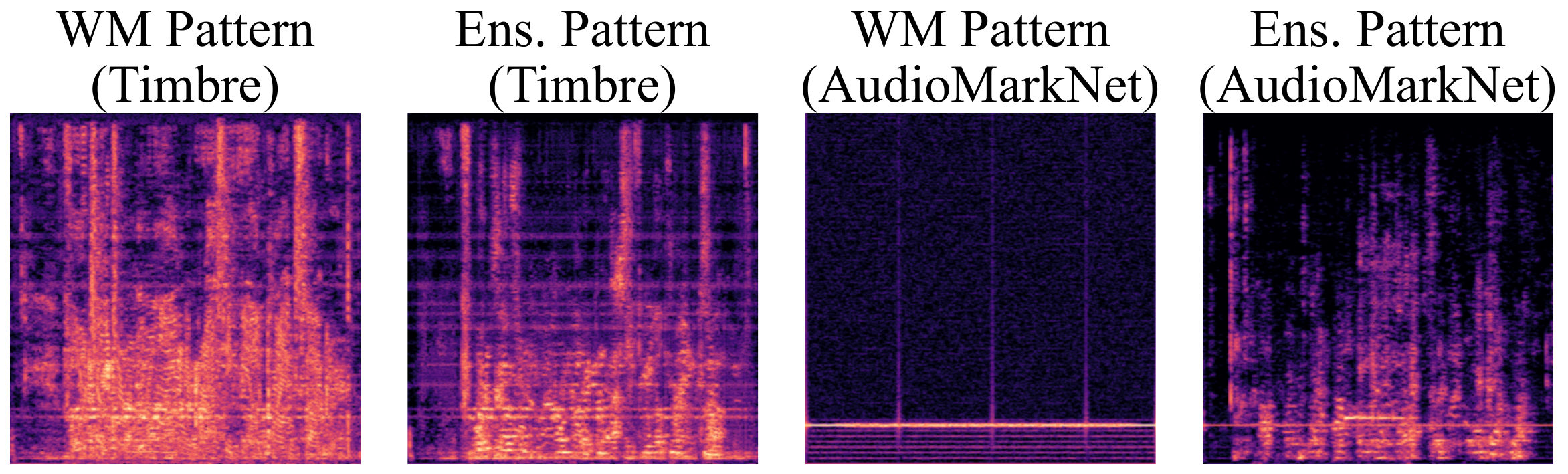}
    \caption{Pattern Visualization. Our ensemble pattern captures some artifacts of the watermark pattern.}
    \label{fig:pattern_visualization_audioseal}
\end{figure}

\subsection{Attack Visualization}

Figure~\ref{fig:audio_visualization_audioseal} compares the spectrograms of the representative audio sample. Square and HSJA introduce strong perturbations into the spectrogram, which mislead the watermark decoder to produce an incorrect output. BigVGAN and DiffErase introduce additional reconstruction artifacts, resulting in perceptible speech distortion.

In contrast, our attack aims to suppress the artifacts that are inconsistent with the speech structure. Specifically, it modifies only the magnitude spectrum while preserving the original phase. By suppressing the non-speech artifacts, our attack successfully removes the watermark while maintaining the speech structure. The red solid boxes highlight regions with noticeable spectrogram distortion produced by existing attacks, whereas the green dashed boxes indicate that our method better preserves the original speech structure.

Figure~\ref{fig:pattern_visualization_audioseal} compares our ensemble pattern and the watermark pattern. We select two representative patterns for Timbre and AudioMarkNet, respectively. The ensemble pattern captures several artifacts present in the watermark pattern, enabling our attack to suppress watermark-related components. Although some low-frequency magnitude components are slightly decreased, our attack still preserves the overall audio quality while effectively removing the watermark. More experiments are shown in Supplementary~\ref{sec:additional_experiment_on_attack_visualization}.

\subsection{Ablation Study}

\begin{figure}[t]
    \centering
    \includegraphics[width=0.95\columnwidth]{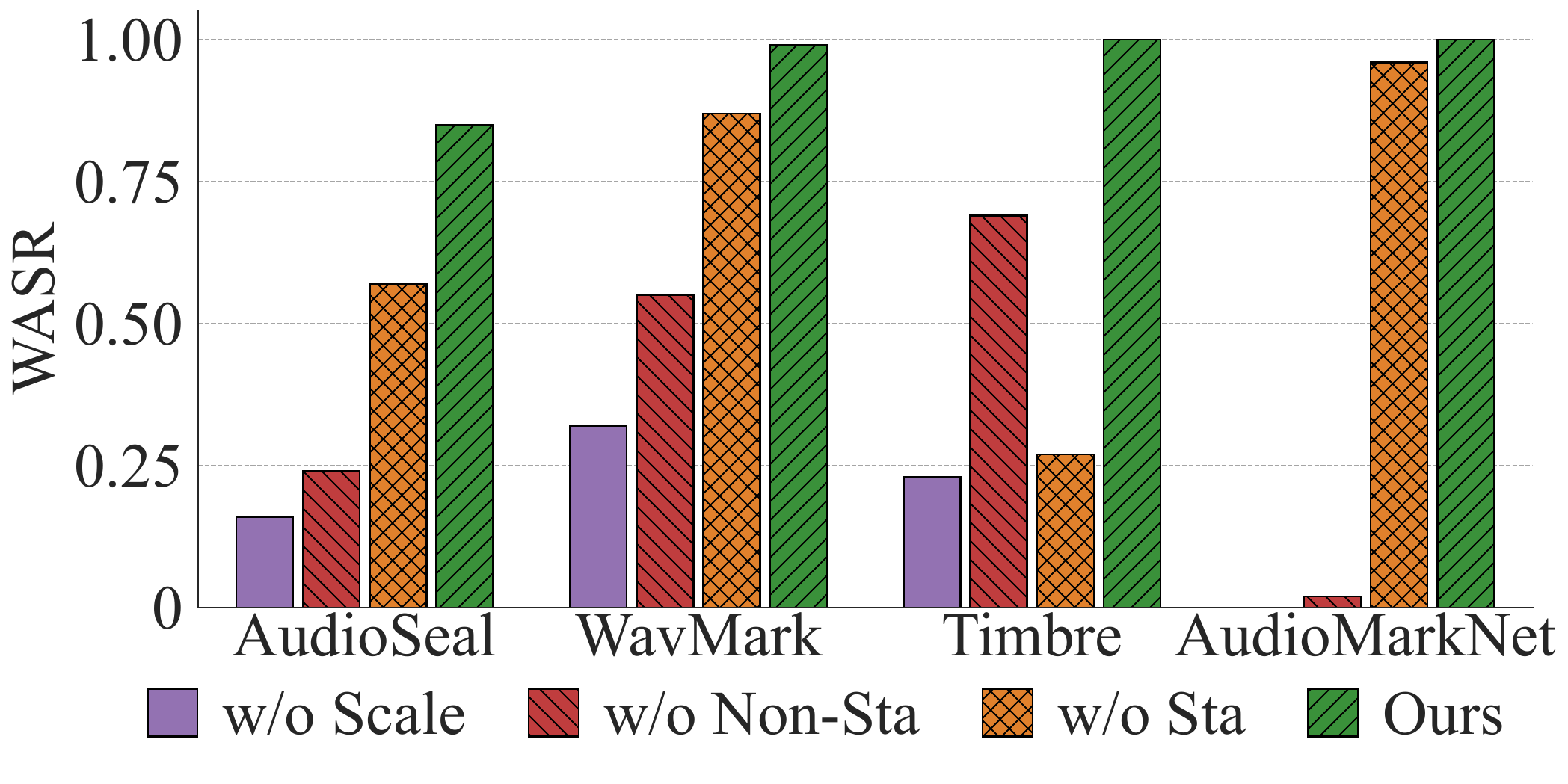}
    \vspace{-2mm}
    \caption{WASR for evaluating adaptive scaling, stationary branch, and non-stationary branch on the WSJ0 dataset.}
    \label{fig:ablation_study_separate}
\end{figure}

We evaluate the effectiveness of adaptive scaling and the two artifact branches. Figure~\ref{fig:ablation_study_separate} shows the WASR results. We set $\alpha=\beta=1$ (w/o Scale), retain only the stationary pattern scaled to 35 dB (w/o Non-sta), and retain only the non-stationary pattern scaled to 20 dB (w/o Sta). Our attack achieves the highest WASR. The two patterns are complementary, and adaptive scaling is necessary for effectively suppressing diverse watermark artifacts. Supplementary~\ref{sec:additional_experiment_on_ablation_study} shows the complete results.

\section{Conclusion}

In this work, we propose \ours, a two-stage query-free black-box attack framework that targets non-speech artifact patterns. Diverse Artifact Learning extracts complementary non-stationary and stationary patterns, and Adaptive Artifact Scaling combines and scales these patterns to construct an ensemble pattern. Moreover, we develop a customized training dataset to facilitate the model-based extraction of stationary patterns. Experimental results demonstrate that \ours achieves a high watermark attack success rate, maintains high perceptual audio quality, and largely preserves linguistic content.


\section*{Limitations}

First, we use empirically chosen target SNRs of 35 dB for the stationary component and 20 dB for the ensemble pattern. Although the scaling coefficients ($\alpha$ and $\beta$) are adaptively adjusted for each sample, these fixed targets may not provide the optimal trade-off between attack effectiveness and perceptual quality for every individual speech recording. Second, the method is designed and evaluated for speech, and its applicability to music and environmental audio remains unexplored.

\section*{Ethical Considerations}

This work studies vulnerabilities in existing audio watermarking methods with the goal of encouraging the development of more robust and secure watermarking systems. We acknowledge that watermark removal attacks can expose weaknesses in current defenses and may potentially facilitate unauthorized voice cloning, impersonation, or evasion of tracing mechanisms. To reduce the risk of misuse, we will not provide or release synthesized speech corresponding to any real individual.

The authors used generative AI tools such as ChatGPT solely for grammar improvement and language polishing. All technical contents were conceived, conducted, and verified by the authors, which include the methodology, experiments, analysis, etc.


\bibliography{custom}

\appendix
\clearpage

\section{Supplementary}

\subsection{Stationary Noise Dataset Generation}

In this paper, the number of selected frequency bins ranges from 10 to 50, their positions are varied, and the SNR ranges from 25 to 35 dB.

\begin{algorithm}
\caption{Construct Stationary Noise Samples}
\label{alg:stationary_noise}
\begin{algorithmic}[1]
\REQUIRE Clean audio waveform $x$, selected frequency bins $\mathcal{F}$, a random SNR value $\rho$, a small constant $\epsilon$
\ENSURE Stationary-noise audio waveform $y_s$
\STATE $S,p \leftarrow \mathrm{STFT}(x)$
\STATE Initialize $S_n(f,l) \leftarrow 0$ for all $f,l$
\FOR{each selected frequency bin $f \in \mathcal{F}$}
\STATE $\bar{S}(f) \leftarrow \frac{1}{L}\sum_{l=1}^{L}S(f,l)$
\STATE $S_n(f,1{:}L) \leftarrow \bar{S}(f)$
\ENDFOR
\STATE $n \leftarrow \mathrm{ISTFT}(S_n,p)$
\STATE $r \leftarrow \sqrt{\frac{\lVert x \rVert_2^2}{\left(\lVert n\rVert_2^2+\epsilon\right) \cdot 10^{\rho/10}}}$
\STATE $n_s \leftarrow r n$
\STATE $y_s \leftarrow x+n_s$
\RETURN $y_s$
\end{algorithmic}
\end{algorithm}

\subsection{Additional Experiment Setup Settings}
\label{sec:additional_experiment_setput_settings}

All experiments are conducted on an NVIDIA RTX 4090 GPU. For AudioSeal, WavMark, and Timbre, we randomly assign a binary watermark message to each audio sample. For AudioMarkNet, we use a fixed binary watermark message. By default, we scale the refined stationary pattern to an SNR of 35 dB and scale the ensemble pattern to an SNR of 20 dB. The sampling rate is set to 16 kHz for all datasets by default.

In our experiments, we ensure that all evaluated audio samples are successfully watermarked. Specifically, we first exclude samples for which watermark embedding is unsuccessful and then use a fixed random seed to select the watermarked samples for evaluation. The experimental settings for the audio watermarking methods are described as follows:
\begin{itemize}
    \item \textbf{AudioSeal, WavMark, and Timbre.} For both Flickr and WSJ0 datasets, we randomly select 100 audio samples using a random seed of 42. The watermark message is randomly generated for each sample; therefore, different audio samples contain different watermark messages.
    \item \textbf{AudioMarkNet.} We use the checkpoint trained on Obama’s voice and adopt the strictest detection criterion: an audio sample is considered watermarked if any of its segments is detected as containing a watermark. Under this criterion, we randomly select 100 audio samples. The watermark message is fixed as \texttt{0000010011000111} for all samples.
\end{itemize}

To train the non-stationary and stationary models and evaluate their performance, we use the VoiceBank-DEMAND dataset as the training set and the Flickr and WSJ0 datasets as inference sets. The training and inference datasets are strictly disjoint. We train both the non-stationary and stationary models using the default parameter configurations provided on the official SGMSE website. For the stationary model, we construct a noisy training dataset according to Algorithm~\ref{alg:stationary_noise}. For TTS-based evaluation and analysis, we follow the settings used in the original Timbre and AudioMarkNet papers. Specifically, Timbre uses the LJSpeech dataset for TTS evaluation, and AudioMarkNet uses the VCTK and Obama datasets. Since the VCTK data used in AudioMarkNet's evaluation overlaps with our training data, we use the Obama dataset for TTS evaluation in our experiments to avoid data overlap.

We categorize the evaluated audio watermark attacks into three types: speech processing attacks, speech enhancement attacks, and adaptive attacks. For speech processing attacks, we use three classical signal processing techniques. For speech enhancement attacks, MP-SENet enhances speech by jointly estimating the magnitude and phase spectra. MetricGAN+ is a classical GAN-based speech enhancement model. StoRM is a diffusion-based speech enhancement model built upon the SGMSE+ architecture and improves denoising performance over the original SGMSE+. For adaptive attacks, BigVGAN employs a neural vocoder for waveform reconstruction. HSJA and Square are query-based attacks that require querying the target watermarking models, we use the default settings provided by AudioMarkBench. DiffErase-Mel and DiffErase-Latent attacks are the query-free attacks. Because BigVGAN, DiffErase-Mel, and DiffErase-Latent are trained on audio sampled at 22.05 kHz, the watermarked audio samples are first resampled from 16 kHz to 22.05 kHz before being fed into these models. The generated audio samples are then resampled back to 16 kHz for evaluation.

\begin{table*}[t]
\centering
{\small
\setlength{\tabcolsep}{0.8mm}
\resizebox{\linewidth}{!}{%
\begin{tabular}{lcccccccccccc}
\toprule
\multirow{2}{*}{\textbf{\shortstack{Attack\\Method}}}
& \multicolumn{3}{c}{\textbf{AudioSeal}}
& \multicolumn{3}{c}{\textbf{WavMark}}
& \multicolumn{3}{c}{\textbf{Timbre}}
& \multicolumn{3}{c}{\textbf{AudioMarkNet}} \\

\cmidrule(lr){2-4}
\cmidrule(lr){5-7}
\cmidrule(lr){8-10}
\cmidrule(lr){11-13}

& \textbf{NISQA}
& \textbf{DNSMOS}
& \textbf{UTMOS}
& \textbf{NISQA}
& \textbf{DNSMOS}
& \textbf{UTMOS}
& \textbf{NISQA}
& \textbf{DNSMOS}
& \textbf{UTMOS}
& \textbf{NISQA}
& \textbf{DNSMOS}
& \textbf{UTMOS} \\
\midrule

w/o attack
& 3.71 & 3.02 & 3.45
& 3.51 & 2.99 & 3.41
& 3.62 & 2.95 & 3.39
& 3.47 & 2.67 & 3.56 \\
\midrule

BigVGAN
& 3.60 & 2.97 & 2.96
& 3.55 & 2.94 & 2.91
& 3.50 & 2.95 & 2.87
& 3.43 & 2.83 & 3.07 \\

Square
& 2.00 & 2.60 & 3.03
& 2.43 & 2.59 & 2.94
& 2.46 & 2.61 & 2.95
& 1.89 & 2.35 & 2.76 \\

HSJA
& 2.58 & 1.94 & 1.65
& 2.86 & 2.66 & 3.10
& 1.31 & 2.05 & 1.83
& 1.07 & 2.01 & 1.58 \\

DiffErase-M
& 3.76 & 3.04 & 2.97
& 3.63 & 3.03 & 2.90
& 3.70 & 3.02 & 2.87
& 3.40 & 2.90 & 3.05 \\

DiffErase-L
& 3.82 & 3.04 & 2.73
& 3.87 & 3.05 & 2.68
& 3.76 & 3.04 & 2.67
& 3.73 & 2.99 & 2.87 \\
\midrule

\textbf{Ours}
& \textbf{\underline{3.98}} & \textbf{\underline{3.08}} & \textbf{\underline{3.51}}
& \textbf{\underline{3.95}} & \textbf{\underline{3.07}} & \textbf{\underline{3.49}}
& \textbf{\underline{3.99}} & \textbf{\underline{3.09}} & \textbf{\underline{3.47}}
& \textbf{\underline{4.21}} & \textbf{\underline{3.09}} & \textbf{\underline{3.70}} \\

\bottomrule

\end{tabular}}
}
\caption{Audio quality evaluation under adaptive attacks on the Flickr dataset. The metrics are NISQA, DNSMOS, and UTMOS.}
\label{tab:audio_quality_flickr}
\end{table*}

\begin{table*}[!t]
\centering

\resizebox{\textwidth}{!}{%
\begin{tabular}{l*{12}{c}}
\toprule
\multirow{3}{*}{\textbf{\shortstack{Attack\\Method}}}
& \multicolumn{12}{c}{\textbf{Flickr dataset}} \\
\cmidrule(lr){2-13}

& \multicolumn{3}{c}{\textbf{AudioSeal}}
& \multicolumn{3}{c}{\textbf{WavMark}}
& \multicolumn{3}{c}{\textbf{Timbre}}
& \multicolumn{3}{c}{\textbf{AudioMarkNet}} \\
\cmidrule(lr){2-4}
\cmidrule(lr){5-7}
\cmidrule(lr){8-10}
\cmidrule(lr){11-13}

& \textbf{SNR} & \textbf{PESQ} & \textbf{ESTOI}
& \textbf{SNR} & \textbf{PESQ} & \textbf{ESTOI}
& \textbf{SNR} & \textbf{PESQ} & \textbf{ESTOI}
& \textbf{SNR} & \textbf{PESQ} & \textbf{ESTOI} \\
\midrule

w/o attack
& 27.44 & 4.44 & 1.00
& 37.67 & 4.04 & 0.99
& 37.79 & 4.10 & 0.99
& 34.69 & 4.05 & 1.00 \\
\midrule

BigVGAN
& -2.53 & 2.91 & 0.91
& -2.54 & 2.72 & 0.90
& -2.53 & 2.72 & 0.90
& -2.64 & 2.75 & 0.90 \\

Square
& 22.30 & 2.30 & 0.96
& 24.17 & 2.31 & 0.96
& 22.66 & 2.27 & 0.95
& 24.82 & 2.42 & 0.86 \\

HSJA
& 6.27 & 1.11 & 0.76
& 24.00 & 2.45 & 0.98
& 9.03 & 1.19 & 0.83
& 4.82 & 1.14 & 0.68 \\

DiffErase-M
& -2.36 & 2.58 & 0.87
& -2.38 & 2.44 & 0.86
& -2.32 & 2.43 & 0.86
& -2.42 & 2.46 & 0.88 \\

DiffErase-L
& -1.71 & 2.18 & 0.81
& -1.62 & 2.10 & 0.81
& -1.66 & 2.15 & 0.81
& -1.69 & 1.99 & 0.79 \\
\midrule

\textbf{Ours}
& \textbf{17.81} & \textbf{3.01} & \textbf{0.93}
& \textbf{18.34} & \textbf{2.85} & \textbf{0.93}
& \textbf{18.11} & \textbf{2.88} & \textbf{0.93}
& \textbf{18.55} & \textbf{3.22} & \textbf{0.87} \\

\bottomrule
\end{tabular}%
}
\caption{Audio quality evaluation on the Flickr dataset under
adaptive attacks. The metrics are SNR, PESQ, and ESTOI.}
\label{tab:audio_quality_flickr_reference_based}
\end{table*}

\begin{table*}[!t]
\centering

\resizebox{\textwidth}{!}{%
\begin{tabular}{l*{12}{c}}
\toprule
\multirow{3}{*}{\textbf{\shortstack{Attack\\Method}}}
& \multicolumn{12}{c}{\textbf{WSJ0 dataset}} \\
\cmidrule(lr){2-13}

& \multicolumn{3}{c}{\textbf{AudioSeal}}
& \multicolumn{3}{c}{\textbf{WavMark}}
& \multicolumn{3}{c}{\textbf{Timbre}}
& \multicolumn{3}{c}{\textbf{AudioMarkNet}} \\
\cmidrule(lr){2-4}
\cmidrule(lr){5-7}
\cmidrule(lr){8-10}
\cmidrule(lr){11-13}

& \textbf{SNR} & \textbf{PESQ} & \textbf{ESTOI}
& \textbf{SNR} & \textbf{PESQ} & \textbf{ESTOI}
& \textbf{SNR} & \textbf{PESQ} & \textbf{ESTOI}
& \textbf{SNR} & \textbf{PESQ} & \textbf{ESTOI} \\
\midrule

w/o attack
& 21.77 & 4.11 & 0.99
& 29.14 & 3.70 & 0.98
& 18.10 & 3.66 & 0.95
& 23.00 & 2.97 & 0.96 \\
\midrule

BigVGAN
& -2.78 & 3.11 & 0.92
& -2.74 & 2.88 & 0.91
& -2.80 & 2.75 & 0.88
& -2.42 & 2.79 & 0.96 \\

Square
& 10.53 & 1.34 & 0.76
& 11.09 & 1.35 & 0.79
& 10.23 & 1.35 & 0.76
& 16.94 & 1.69 & 0.87 \\

HSJA
& 3.60 & 1.12 & 0.67
& 12.49 & 1.84 & 0.91
& 0.18 & 1.10 & 0.62
& -6.24 & 1.08 & 0.45 \\

DiffErase-M
& -2.45 & 2.72 & 0.88
& -2.42 & 2.59 & 0.87
& -2.52 & 2.50 & 0.85
& -2.49 & 2.46 & 0.86 \\

DiffErase-L
& -2.12 & 2.50 & 0.82
& -2.10 & 2.35 & 0.81
& -2.21 & 2.37 & 0.81
& -2.06 & 2.12 & 0.79 \\
\midrule

\textbf{Ours}
& \textbf{18.89} & \textbf{3.35} & \textbf{0.96}
& \textbf{19.96} & \textbf{2.91} & \textbf{0.94}
& \textbf{16.39} & \textbf{2.94} & \textbf{0.92}
& \textbf{22.46} & \textbf{3.57} & \textbf{0.97} \\

\bottomrule
\end{tabular}%
}
\caption{Audio quality evaluation on the WSJ0 dataset under
adaptive attacks. The metrics are SNR, PESQ, and ESTOI.}
\label{tab:audio_quality_wsj0_reference_based}
\end{table*}

\begin{table*}[t]
\centering
\setlength{\tabcolsep}{0.8mm}
\renewcommand{\arraystretch}{0.95}

\resizebox{\linewidth}{!}{%
\begin{tabular}{ll*{16}{c}}
\toprule
\multirow{3}{*}{ASR Model}
& \multirow{3}{*}{Attack Method}
& \multicolumn{8}{c}{\textbf{Flickr dataset}}
& \multicolumn{8}{c}{\textbf{WSJ0 dataset}} \\

\cmidrule(lr){3-10}
\cmidrule(lr){11-18}

& & \multicolumn{2}{c}{AudioSeal}
& \multicolumn{2}{c}{WavMark}
& \multicolumn{2}{c}{Timbre}
& \multicolumn{2}{c}{AudioMarkNet}
& \multicolumn{2}{c}{AudioSeal}
& \multicolumn{2}{c}{WavMark}
& \multicolumn{2}{c}{Timbre}
& \multicolumn{2}{c}{AudioMarkNet} \\

\cmidrule(lr){3-4}
\cmidrule(lr){5-6}
\cmidrule(lr){7-8}
\cmidrule(lr){9-10}
\cmidrule(lr){11-12}
\cmidrule(lr){13-14}
\cmidrule(lr){15-16}
\cmidrule(lr){17-18}

& & WER & CER & WER & CER & WER & CER & WER & CER
    & WER & CER & WER & CER & WER & CER & WER & CER \\
\midrule

\multirow{14}{*}{Whisper}
& Clean
& 5.36 & 1.95 & 5.36 & 1.95 & 5.36 & 1.95 & 5.52 & 2.54
& 8.41 & 5.60 & 8.41 & 5.60 & 8.41 & 5.60 & 3.86 & 2.66 \\

& Watermark
& 5.62 & 2.06 & 5.45 & 1.99 & 5.96 & 2.26 & 5.89 & 3.03
& 8.58 & 5.69 & 8.52 & 5.72 & 8.35 & 5.58 & 3.86 & 2.66 \\

& NoiseReduce
& 8.21 & 3.49 & 7.35 & 3.04 & 7.87 & 3.37 & 7.09 & 3.35
& 9.91 & 6.36 & 10.03 & 6.50 & 9.74 & 6.10 & 5.54 & 3.34 \\

& Wiener Filter
& 5.79 & 2.16 & 6.22 & 2.78 & 6.14 & 2.35 & 5.71 & 2.35
& 9.39 & 6.16 & 9.10 & 5.96 & 8.52 & 5.76 & 4.58 & 2.78 \\

& MMSE\_STSA
& 5.53 & 2.00 & 5.19 & 1.90 & 5.88 & 2.28 & 5.34 & 2.41
& 8.64 & 5.70 & 8.93 & 5.93 & 8.17 & 5.58 & 4.10 & 2.56 \\

& MP-SENet
& 6.05 & 2.54 & 6.40 & 2.63 & 6.31 & 2.42 & 7.18 & 3.63
& 8.87 & 6.25 & 9.10 & 6.26 & 8.64 & 6.02 & 3.86 & 2.50 \\

& MetricGAN+
& 8.47 & 3.54 & 7.43 & 3.30 & 7.95 & 3.64 & 7.46 & 3.69
& 14.78 & 8.56 & 15.01 & 8.98 & 16.23 & 9.44 & 5.54 & 3.40 \\

& StoRM
& 6.31 & 2.80 & 8.47 & 3.89 & 6.57 & 2.44 & 7.92 & 4.08
& 8.58 & 5.69 & 8.87 & 5.77 & 8.41 & 5.66 & 4.10 & 2.66 \\

& BigVGAN
& 6.48 & 2.37 & 6.14 & 2.30 & 5.96 & 2.23 & 5.89 & 2.71
& 8.93 & 5.71 & 9.22 & 5.92 & 8.52 & 5.69 & 4.58 & 3.02 \\

& Square
& 7.69 & 3.70 & 7.35 & 3.21 & 7.78 & 3.56 & 6.63 & 2.84
& 16.35 & 9.91 & 14.38 & 8.70 & 16.12 & 9.80 & 6.27 & 3.28 \\

& HSJA
& 13.22 & 7.06 & 5.79 & 2.11 & 9.94 & 4.65 & 14.09 & 8.41
& 16.17 & 10.19 & 9.39 & 6.01 & 20.70 & 13.31 & 48.55 & 37.42 \\

& DiffErase-M
& 8.04 & 3.99 & 8.12 & 4.06 & 8.30 & 4.27 & 8.66 & 4.55
& 8.64 & 5.64 & 9.16 & 5.88 & 8.64 & 5.68 & 4.46 & 2.84 \\

& DiffErase-L
& 7.35 & 2.75 & 11.67 & 6.67 & 8.39 & 3.58 & 6.63 & 3.09
& 9.16 & 5.95 & 9.57 & 6.09 & 9.22 & 5.97 & 4.58 & 2.96 \\

& \textbf{Ours}
& \textbf{7.69} & \textbf{3.26}
& \textbf{7.61} & \textbf{3.11}
& \textbf{7.43} & \textbf{3.30}
& \textbf{9.12} & \textbf{4.99}
& \textbf{8.75} & \textbf{5.75}
& \textbf{9.04} & \textbf{5.95}
& \textbf{8.06} & \textbf{5.62}
& \textbf{4.22} & \textbf{2.78} \\

\midrule

\multirow{14}{*}{wav2vec}
& Clean
& 10.46 & 2.78 & 10.46 & 2.78 & 10.46 & 2.78 & 8.84 & 3.07
& 14.43 & 6.21 & 14.43 & 6.21 & 14.43 & 6.21 & 11.45 & 3.48 \\

& Watermark
& 10.46 & 2.80 & 11.24 & 3.18 & 10.72 & 2.94 & 8.56 & 3.05
& 14.49 & 6.25 & 15.07 & 6.39 & 14.78 & 6.21 & 12.29 & 3.64 \\

& NoiseReduce
& 13.66 & 4.59 & 13.05 & 4.61 & 13.83 & 4.94 & 14.18 & 5.59
& 15.88 & 6.77 & 15.65 & 6.69 & 16.06 & 6.86 & 13.37 & 4.31 \\

& Wiener Filter
& 11.24 & 3.28 & 11.41 & 3.51 & 11.06 & 3.23 & 9.85 & 3.40
& 14.96 & 6.44 & 15.42 & 6.49 & 15.07 & 6.40 & 12.65 & 3.78 \\

& MMSE\_STSA
& 10.03 & 2.69 & 10.72 & 3.06 & 10.37 & 3.02 & 8.93 & 3.22
& 14.90 & 6.23 & 15.42 & 6.27 & 14.84 & 6.29 & 12.17 & 3.50 \\

& MP-SENet
& 11.32 & 3.77 & 12.01 & 3.96 & 10.72 & 3.14 & 9.67 & 3.48
& 14.67 & 6.72 & 15.48 & 7.03 & 15.42 & 6.91 & 11.69 & 3.42 \\

& MetricGAN+
& 12.01 & 4.01 & 12.71 & 4.34 & 13.14 & 4.42 & 11.14 & 4.82
& 21.80 & 9.21 & 22.78 & 9.67 & 22.26 & 9.58 & 14.34 & 4.63 \\

& StoRM
& 13.40 & 4.72 & 14.17 & 4.77 & 13.74 & 4.72 & 11.51 & 4.59
& 14.61 & 6.19 & 15.07 & 6.47 & 15.30 & 6.45 & 11.33 & 3.50 \\

& BigVGAN
& 11.75 & 3.49 & 13.31 & 4.09 & 12.36 & 3.75 & 12.06 & 5.00
& 15.94 & 6.72 & 15.54 & 6.72 & 15.77 & 6.74 & 12.53 & 3.62 \\

& Square
& 13.48 & 4.63 & 13.48 & 4.77 & 13.92 & 4.63 & 12.71 & 5.76
& 27.19 & 11.43 & 25.74 & 10.80 & 26.90 & 11.43 & 20.72 & 8.73 \\

& HSJA
& 36.82 & 20.21 & 11.41 & 3.51 & 22.56 & 9.46 & 40.06 & 22.82
& 53.45 & 28.08 & 15.94 & 6.63 & 60.64 & 31.97 & 84.70 & 72.59 \\

& DiffErase-M
& 14.78 & 6.00 & 15.90 & 6.51 & 14.43 & 5.87 & 13.54 & 5.83
& 15.71 & 6.82 & 16.58 & 7.06 & 16.35 & 6.91 & 12.65 & 3.86 \\

& DiffErase-L
& 13.92 & 4.40 & 15.56 & 5.20 & 13.92 & 4.40 & 13.54 & 5.44
& 15.59 & 6.73 & 15.54 & 6.73 & 16.70 & 6.91 & 12.53 & 3.88 \\

& \textbf{Ours}
& \textbf{12.88} & \textbf{4.39}
& \textbf{13.92} & \textbf{4.96}
& \textbf{13.74} & \textbf{4.87}
& \textbf{13.35} & \textbf{5.70}
& \textbf{14.38} & \textbf{6.16}
& \textbf{14.49} & \textbf{6.34}
& \textbf{15.48} & \textbf{6.63}
& \textbf{11.69} & \textbf{3.42} \\

\midrule

\multirow{14}{*}{SpeechBrain}
& Clean
& 13.57 & 6.89 & 13.57 & 6.89 & 13.57 & 6.89 & 10.22 & 5.40
& 14.84 & 7.67 & 14.84 & 7.67 & 14.84 & 7.67 & 14.94 & 7.89 \\

& Watermark
& 13.74 & 7.18 & 13.92 & 7.70 & 13.40 & 7.06 & 10.41 & 5.49
& 15.77 & 8.44 & 17.33 & 9.55 & 17.74 & 9.96 & 15.42 & 7.97 \\

& NoiseReduce
& 14.17 & 5.87 & 14.95 & 6.94 & 15.21 & 7.20 & 10.41 & 5.49
& 24.12 & 16.03 & 20.35 & 12.00 & 21.33 & 12.58 & 17.35 & 8.77 \\

& Wiener Filter
& 13.05 & 6.23 & 14.87 & 8.08 & 15.82 & 8.84 & 9.85 & 5.44
& 18.26 & 10.53 & 23.13 & 14.60 & 22.20 & 14.39 & 15.42 & 7.67 \\

& MMSE\_STSA
& 14.43 & 7.36 & 13.66 & 6.82 & 17.29 & 10.85 & 10.68 & 5.68
& 17.04 & 9.68 & 17.04 & 9.44 & 17.68 & 9.99 & 11.69 & 4.21 \\

& MP-SENet
& 15.13 & 7.74 & 14.26 & 6.80 & 15.73 & 9.17 & 11.69 & 6.11
& 16.87 & 9.99 & 16.64 & 8.97 & 17.51 & 9.94 & 13.98 & 6.45 \\

& MetricGAN+
& 15.73 & 9.22 & 16.68 & 9.41 & 21.26 & 15.25 & 11.60 & 5.96
& 29.57 & 17.13 & 31.42 & 17.91 & 31.65 & 18.72 & 17.11 & 8.25 \\

& StoRM
& 16.16 & 7.79 & 16.34 & 7.67 & 16.34 & 8.03 & 14.55 & 8.86
& 17.68 & 10.61 & 18.26 & 10.66 & 16.64 & 9.15 & 13.37 & 6.25 \\

& BigVGAN
& 12.96 & 6.53 & 14.09 & 9.17 & 12.79 & 5.51 & 11.97 & 7.22
& 16.93 & 8.59 & 16.93 & 8.95 & 21.80 & 13.59 & 15.54 & 7.69 \\

& Square
& 18.67 & 11.76 & 14.95 & 7.98 & 16.25 & 9.47 & 14.64 & 8.95
& 27.07 & 17.43 & 22.79 & 14.01 & 28.46 & 18.23 & 18.43 & 8.27 \\

& HSJA
& 27.66 & 16.27 & 17.29 & 10.31 & 21.95 & 12.99 & 19.24 & 11.64
& 28.99 & 18.99 & 16.70 & 8.60 & 34.78 & 21.38 & 51.93 & 35.70 \\

& DiffErase-M
& 23.77 & 18.51 & 17.55 & 10.33 & 17.46 & 9.88 & 14.83 & 9.12
& 21.16 & 13.69 & 24.52 & 15.53 & 23.94 & 15.58 & 15.42 & 6.91 \\

& DiffErase-L
& 16.85 & 10.66 & 20.57 & 13.56 & 17.11 & 9.60 & 13.26 & 7.94
& 18.49 & 10.19 & 24.64 & 16.24 & 24.12 & 17.31 & 15.66 & 7.69 \\

& \textbf{Ours}
& \textbf{16.77} & \textbf{8.88}
& \textbf{17.46} & \textbf{9.84}
& \textbf{18.24} & \textbf{11.35}
& \textbf{15.19} & \textbf{9.12}
& \textbf{17.45} & \textbf{9.84}
& \textbf{20.35} & \textbf{13.19}
& \textbf{22.49} & \textbf{13.98}
& \textbf{14.10} & \textbf{6.59} \\

\bottomrule
\end{tabular}%
}
\caption{WER (\%) and CER (\%) of different attack methods evaluated on three ASR models.}
\label{tab:asr_results_flickr_wsj0}
\end{table*}

\subsection{Additional Experiment on Audio Quality}
\label{sec:additional_experiment_on_audio_quality}

In the main paper, we report the audio quality results on the WSJ0 dataset. Here, we provide the corresponding results on the Flickr dataset. As shown in Table~\ref{tab:audio_quality_flickr}, our method consistently achieves the highest NISQA, DNSMOS, and UTMOS scores among all evaluated attacks across the four watermarking methods, demonstrating its ability to preserve perceptual audio quality while removing watermarks.

The main paper mainly uses three MOS-based metrics (NISQA, DNSMOS, and UTMOS) to evaluate perceptual audio quality. To provide a more comprehensive evaluation, we additionally report three reference-based metrics, SNR, PESQ, and ESTOI. We use the corresponding clean audio as the reference. The reconstructed audio may differ slightly in length from the original audio; therefore, we truncate each reconstructed–clean pair to their common minimum length before evaluation. Table~\ref{tab:audio_quality_flickr_reference_based} and Table~\ref{tab:audio_quality_wsj0_reference_based} present the results on the Flickr and WSJ0 datasets. The results show that BigVGAN and DiffErase yield negative average SNR values despite achieving relatively high WASRs against several watermarking methods. Square achieves relatively high SNR and, in some cases, high ESTOI, but obtains lower PESQ and MOS-based scores. Moreover, high scores on individual quality metrics do not correspond to successful watermark removal. HSJA achieves a relatively high WASR but substantially degrades audio quality. In comparison, our method achieves stronger overall performance across the three reference-based metrics while providing a better balance between attack effectiveness and audio quality. In particular, our method achieves higher average PESQ and ESTOI scores than the other five attacks.

\begin{table*}[t]
    \centering
    \setlength{\tabcolsep}{2pt}
    \resizebox{\textwidth}{!}{%
    \begin{tabular}{cl*{12}{c}}
        \toprule
        \multirow{2}{*}{TTS model}
        & \multirow{2}{*}{Attack method}
        & \multicolumn{6}{c}{Timbre}
        & \multicolumn{6}{c}{AudioMarkNet} \\
        \cmidrule(lr){3-8}
        \cmidrule(lr){9-14}
        & & WASR & NISQA & UTMOS & DNSMOS & WER & CER
          & WASR & NISQA & UTMOS & DNSMOS & WER & CER \\
        \midrule

        \multirow{7}{*}{\shortstack{FastSpeech2\\+HiFi-GAN}}
        & Clean
        & 1.00 & 3.59 & 4.10 & 3.22 & 6.97 & 3.26
        & 0.93 & 2.82 & 2.30 & 2.58 & 21.13 & 10.35 \\

        & Watermark
        & 0.00 & 4.58 & 4.13 & 3.24 & 5.96 & 2.77
        & 0.00 & 2.97 & 2.29 & 2.55 & 22.44 & 11.52 \\

        & Square
        & 0.92 & 2.33 & 3.45 & 2.82 & 6.50 & 2.97
        & 0.02 & 2.39 & 2.03 & 2.37 & 31.28 & 16.41 \\

        & HSJA
        & 0.95 & 2.33 & 3.45 & 2.81 & 6.56 & 3.18
        & 0.76 & 2.33 & 1.24 & 2.00 & 54.01 & 31.10 \\

        & DiffErase-M
        & 1.00 & 4.48 & 3.99 & 3.19 & 6.85 & 3.07
        & 0.09 & 2.85 & 2.44 & 2.66 & 20.77 & 9.86 \\

        & DiffErase-L
        & 1.00 & 4.26 & 3.83 & 3.20 & 6.67 & 3.17
        & 0.57 & 3.19 & 2.23 & 2.64 & 19.21 & 8.64 \\

        & \textbf{Ours}
        & \textbf{1.00} & \textbf{4.84} & \textbf{4.12}
        & \textbf{3.12} & \textbf{6.20} & \textbf{2.90}
        & \textbf{0.99} & \textbf{3.78} & \textbf{2.75}
        & \textbf{2.74} & \textbf{19.87} & \textbf{9.83} \\
        \midrule

        \multirow{7}{*}{\shortstack{Tacotron2\\+HiFi-GAN}}
        & Clean
        & 1.00 & 4.17 & 4.28 & 3.33 & 5.24 & 2.64
        & 0.91 & 2.74 & 2.01 & 2.58 & 20.38 & 11.04 \\

        & Watermark
        & 0.00 & 4.49 & 3.86 & 3.25 & 6.08 & 2.95
        & 0.01 & 2.86 & 2.06 & 2.59 & 20.73 & 11.11 \\

        & Square
        & 1.00 & 1.96 & 3.01 & 2.68 & 6.85 & 3.58
        & 0.01 & 2.36 & 1.80 & 2.48 & 23.54 & 16.65 \\

        & HSJA
        & 1.00 & 2.11 & 3.19 & 2.72 & 6.20 & 3.20
        & 0.39 & 0.68 & 1.27 & 1.87 & 83.51 & 60.77 \\

        & DiffErase-M
        & 1.00 & 4.54 & 4.00 & 3.26 & 6.38 & 3.06
        & 0.02 & 2.73 & 2.08 & 2.77 & 24.26 & 12.40 \\

        & DiffErase-L
        & 1.00 & 4.30 & 3.63 & 3.16 & 5.48 & 2.87
        & 0.53 & 2.74 & 1.69 & 2.54 & 21.81 & 11.12 \\

        & \textbf{Ours}
        & \textbf{1.00} & \textbf{4.87} & \textbf{4.11}
        & \textbf{3.28} & \textbf{5.42} & \textbf{2.66}
        & \textbf{1.00} & \textbf{3.86} & \textbf{2.53}
        & \textbf{2.70} & \textbf{19.90} & \textbf{10.48} \\
        \midrule

        \multirow{7}{*}{YourTTS}
        & Clean
        & 1.00 & 4.27 & 3.60 & 3.14 & 18.06 & 8.82
        & 0.96 & 3.33 & 2.69 & 2.79 & 43.39 & 23.68 \\

        & Watermark
        & 0.02 & 4.52 & 3.48 & 3.11 & 18.12 & 8.83
        & 0.00 & 3.42 & 2.69 & 2.73 & 43.15 & 23.28 \\

        & Square
        & 0.95 & 2.19 & 2.81 & 2.82 & 32.42 & 17.00
        & 0.01 & 2.44 & 2.44 & 2.73 & 47.62 & 27.09 \\

        & HSJA
        & 0.88 & 2.21 & 2.86 & 2.82 & 30.39 & 16.20
        & 0.17 & 0.73 & 1.28 & 2.19 & 110.00 & 80.91 \\

        & DiffErase-M
        & 1.00 & 4.51 & 3.10 & 2.92 & 25.45 & 12.36
        & 0.05 & 3.15 & 2.28 & 2.92 & 44.70 & 25.07 \\

        & DiffErase-L
        & 1.00 & 4.16 & 3.07 & 3.01 & 24.43 & 12.24
        & 0.12 & 3.25 & 2.63 & 2.78 & 43.27 & 22.60 \\

        & \textbf{Ours}
        & \textbf{1.00} & \textbf{4.92} & \textbf{3.67}
        & \textbf{3.19} & \textbf{17.58} & \textbf{8.37}
        & \textbf{0.99} & \textbf{4.37} & \textbf{3.09}
        & \textbf{3.14} & \textbf{38.08} & \textbf{19.98} \\

        \bottomrule
    \end{tabular}%
    }
    \caption{TTS synthesis performance evaluation. Whisper outputs the WER and CER.}
    \label{tab:tts_parts_complete_results}
\end{table*}

\subsection{Additional Experiment on Automatic Speech Recognition Evaluation}
\label{sec:additional_experiment_on_automatic_speech_recognition_evaluation}

Table~\ref{tab:asr_results_flickr_wsj0} shows the complete WER and CER results for three automatic speech recognition models evaluated on the Flickr dataset and the WSJ0 dataset. Compared with other attacks, our method achieves strong attack effectiveness while maintaining relatively low WER and CER, indicating better preservation of the linguistic content.

\subsection{Additional Experiment on Text-to-Speech Analysis}
\label{sec:addtional_experiment_on_text_to_speech_analysis}

Table~\ref{tab:tts_parts_complete_results} shows the complete results for evaluating the subsequent TTS-synthesized speech. We use Whisper Base to compute WER and CER. For each attack, we fine-tune the TTS model on the attacked watermarked samples and generate 100 utterances for evaluation. Clean and Watermark denote fine-tuning on clean and unmodified watermarked samples, respectively.

To generate the 100 evaluation utterances, we use the predefined test file list provided by Tacotron 2 and randomly select 100 utterances using a fixed random seed of 42. The utterances in this test set are excluded from the fine-tuning process. For dataset selection, we follow the settings used in the original Timbre and AudioMarkNet papers. Timbre uses LJSpeech samples for fine-tuning, AudioMarkNet uses the Obama speech samples provided in the original paper. The data are split into 90\% training and 10\% testing sets. The fine-tuning configurations for the three TTS models are as follows:

\begin{itemize}
    \item FastSpeech 2: For Timbre, we use a batch size of 16. For AudioMarkNet, we use a batch size of 4. Both use 10,000 fine-tuning steps and a learning rate of 5e-5. The initial checkpoint is the LibriTTS checkpoint downloaded from the FastSpeech 2 website. All other configurations follow the default settings. 
    \item Tacotron 2: For Timbre, we use a batch size of 16 and 2,600 fine-tuning steps. For AudioMarkNet, we use a batch size of 4 and 1,000 fine-tuning steps. The initial checkpoint is downloaded from the Tacotron 2 website. All other configurations follow the default settings.
    \item YourTTS: For Timbre, we use a batch size of 32 and fine-tune for 10 epochs. For AudioMarkNet, we use a batch size of 4 and fine-tune for 30 epochs. Both use a learning rate of 2e-5. The initial checkpoint is the VCTK checkpoint downloaded from the YourTTS website. All other configurations follow the default settings.
\end{itemize}

\begin{table*}[t]
    \centering
    \setlength{\tabcolsep}{3pt}
    \resizebox{\textwidth}{!}{%
    \begin{tabular}{l*{4}{cccc}}
        \toprule
        \multirow{2}{*}{Attack Method}
        & \multicolumn{4}{c}{AudioSeal}
        & \multicolumn{4}{c}{WavMark}
        & \multicolumn{4}{c}{Timbre}
        & \multicolumn{4}{c}{AudioMarkNet} \\
        \cmidrule(lr){2-5}
        \cmidrule(lr){6-9}
        \cmidrule(lr){10-13}
        \cmidrule(lr){14-17}
        & WASR & NISQA & DNSMOS & UTMOS
        & WASR & NISQA & DNSMOS & UTMOS
        & WASR & NISQA & DNSMOS & UTMOS
        & WASR & NISQA & DNSMOS & UTMOS \\
        \midrule

        w/o scale
        & 0.12 & 4.26 & 3.13 & 3.56
        & 0.25 & 4.27 & 3.15 & 3.54
        & 0.57 & 4.27 & 3.13 & 3.52
        & 0.00 & 4.41 & 3.11 & 3.75 \\

        w/o Non-Sta
        & 0.71 & 3.38 & 2.88 & 3.37
        & 0.61 & 3.54 & 2.94 & 3.35
        & 0.85 & 3.59 & 2.91 & 3.35
        & 0.86 & 3.68 & 2.92 & 3.59 \\

        w/o Sta
        & 0.21 & 4.21 & 3.14 & 3.56
        & 0.48 & 4.14 & 3.14 & 3.53
        & 0.31 & 4.22 & 3.15 & 3.52
        & 0.93 & 4.29 & 3.11 & 3.74 \\

        \textbf{Ours}
        & 0.84 & 3.98 & 3.08 & 3.51
        & 0.91 & 3.95 & 3.07 & 3.49
        & 0.94 & 3.99 & 3.09 & 3.47
        & 1.00 & 4.21 & 3.09 & 3.70 \\

        \bottomrule
    \end{tabular}%
    }
    \caption{Ablation results on the Flickr dataset.}
    \label{tab:ablation_flickr}
\end{table*}

\begin{table*}[t]
    \centering
    \setlength{\tabcolsep}{3pt}
    \resizebox{\textwidth}{!}{%
    \begin{tabular}{l*{4}{cccc}}
        \toprule
        \multirow{2}{*}{Attack Method}
        & \multicolumn{4}{c}{AudioSeal}
        & \multicolumn{4}{c}{WavMark}
        & \multicolumn{4}{c}{Timbre}
        & \multicolumn{4}{c}{AudioMarkNet} \\
        \cmidrule(lr){2-5}
        \cmidrule(lr){6-9}
        \cmidrule(lr){10-13}
        \cmidrule(lr){14-17}
        & WASR & NISQA & DNSMOS & UTMOS
        & WASR & NISQA & DNSMOS & UTMOS
        & WASR & NISQA & DNSMOS & UTMOS
        & WASR & NISQA & DNSMOS & UTMOS \\
        \midrule

        w/o scale
        & 0.16 & 4.24 & 3.30 & 4.31
        & 0.32 & 4.38 & 3.32 & 4.27
        & 0.23 & 4.24 & 3.29 & 4.20
        & 0.00 & 4.51 & 3.35 & 4.25 \\

        w/o Non-Sta
        & 0.24 & 3.73 & 3.18 & 4.15
        & 0.55 & 4.11 & 3.25 & 4.11
        & 0.69 & 3.76 & 3.20 & 4.05
        & 0.02 & 4.04 & 3.14 & 4.16 \\

        w/o Sta
        & 0.57 & 4.16 & 3.27 & 4.28
        & 0.87 & 4.22 & 3.26 & 4.19
        & 0.27 & 4.22 & 3.28 & 4.18
        & 0.96 & 4.58 & 3.32 & 4.23 \\

        \textbf{Ours}
        & 0.85 & 4.22 & 3.25 & 4.25
        & 0.99 & 4.29 & 3.22 & 4.18
        & 1.00 & 4.28 & 3.23 & 4.10
        & 1.00 & 4.59 & 3.31 & 4.22 \\

        \bottomrule
    \end{tabular}%
    }
    \caption{Ablation results on the WSJ0 dataset.}
    \label{tab:ablation_wsj0}
\end{table*}

\subsection{Additional Experiment on Attack Visualization}
\label{sec:additional_experiment_on_attack_visualization}

\begin{figure*}[t]
    \centering
    \includegraphics[width=0.75\linewidth]{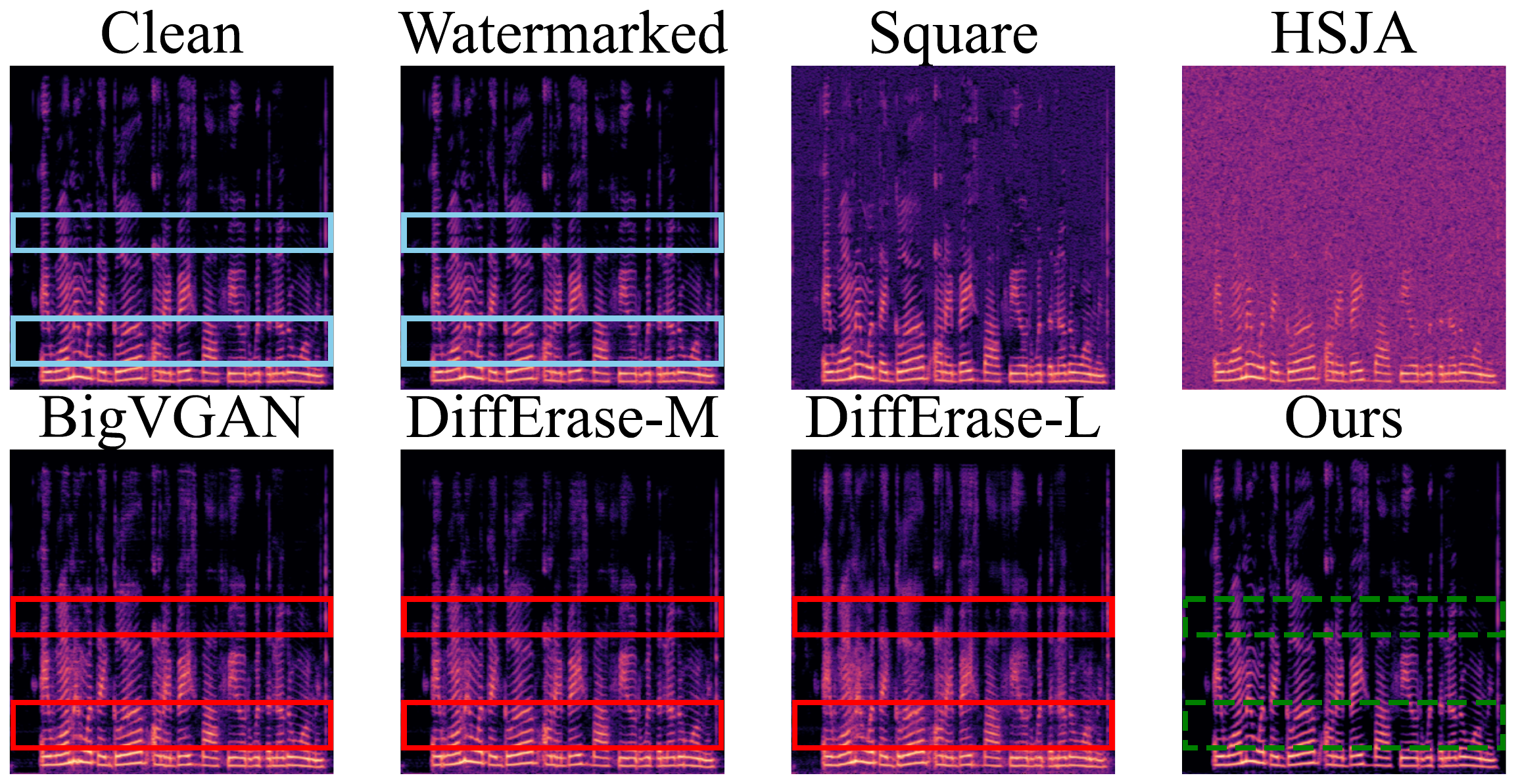}
    \caption{Attacked audio visualization (AudioSeal).}
    \label{fig:audio_visualization_audioseal_sup}
\end{figure*}

\begin{figure*}[t]
    \centering
    \includegraphics[width=\linewidth]{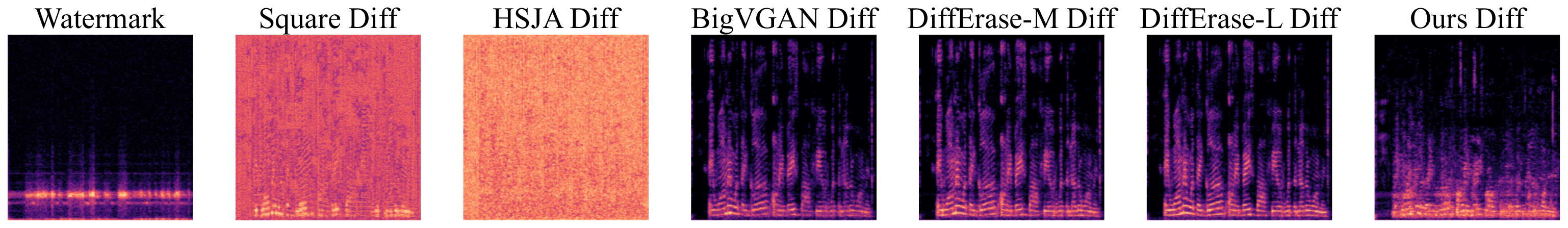}
    \caption{Pattern Visualization (AudioSeal).}
    \label{fig:audio_visualization_audioseal_pattern_difference}
\end{figure*}

\begin{figure*}[t]
    \centering
    \includegraphics[width=0.75\linewidth]{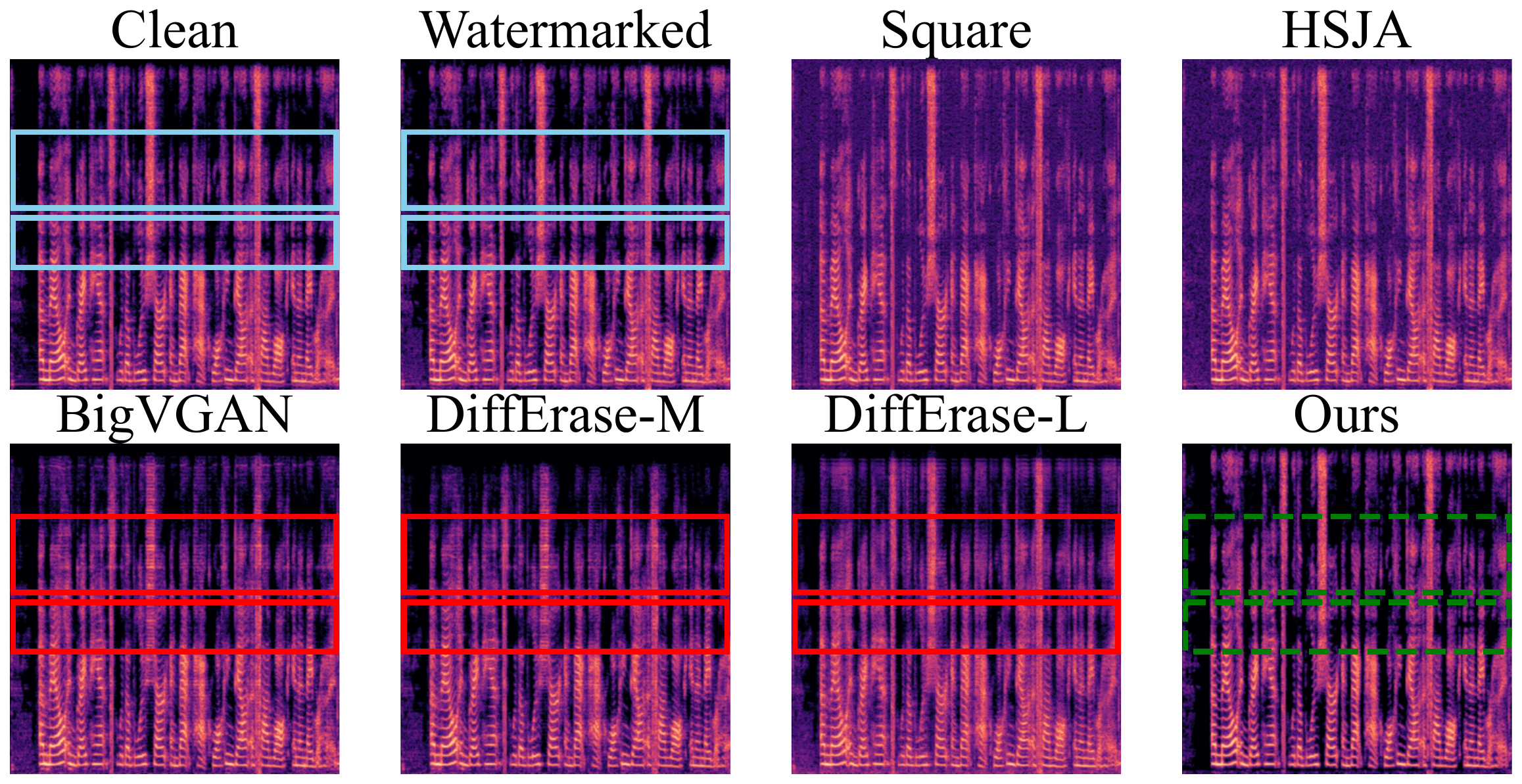}
    \caption{Attacked audio visualization (WavMark).}
    \label{fig:audio_visualization_wavmark}
\end{figure*}

\begin{figure*}[t]
    \centering
    \includegraphics[width=\linewidth]{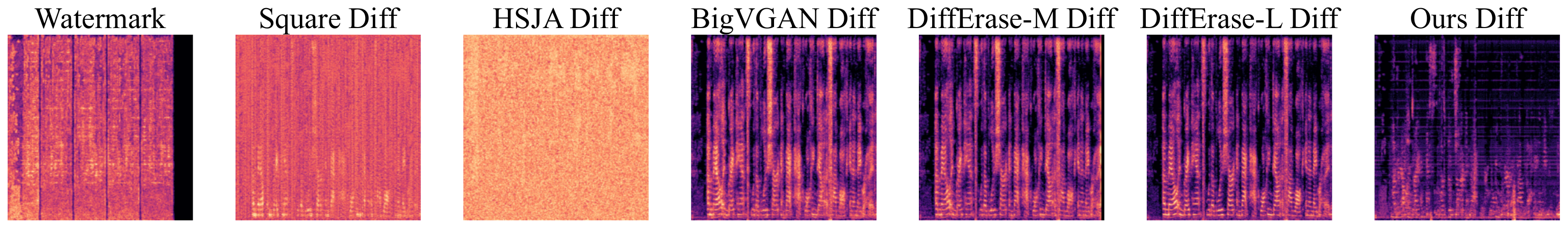}
    \caption{Pattern Visualization (WavMark).}
    \label{fig:audio_visualization_wavmark_pattern_difference}
\end{figure*}

\begin{figure*}[t]
    \centering
    \includegraphics[width=0.75\linewidth]{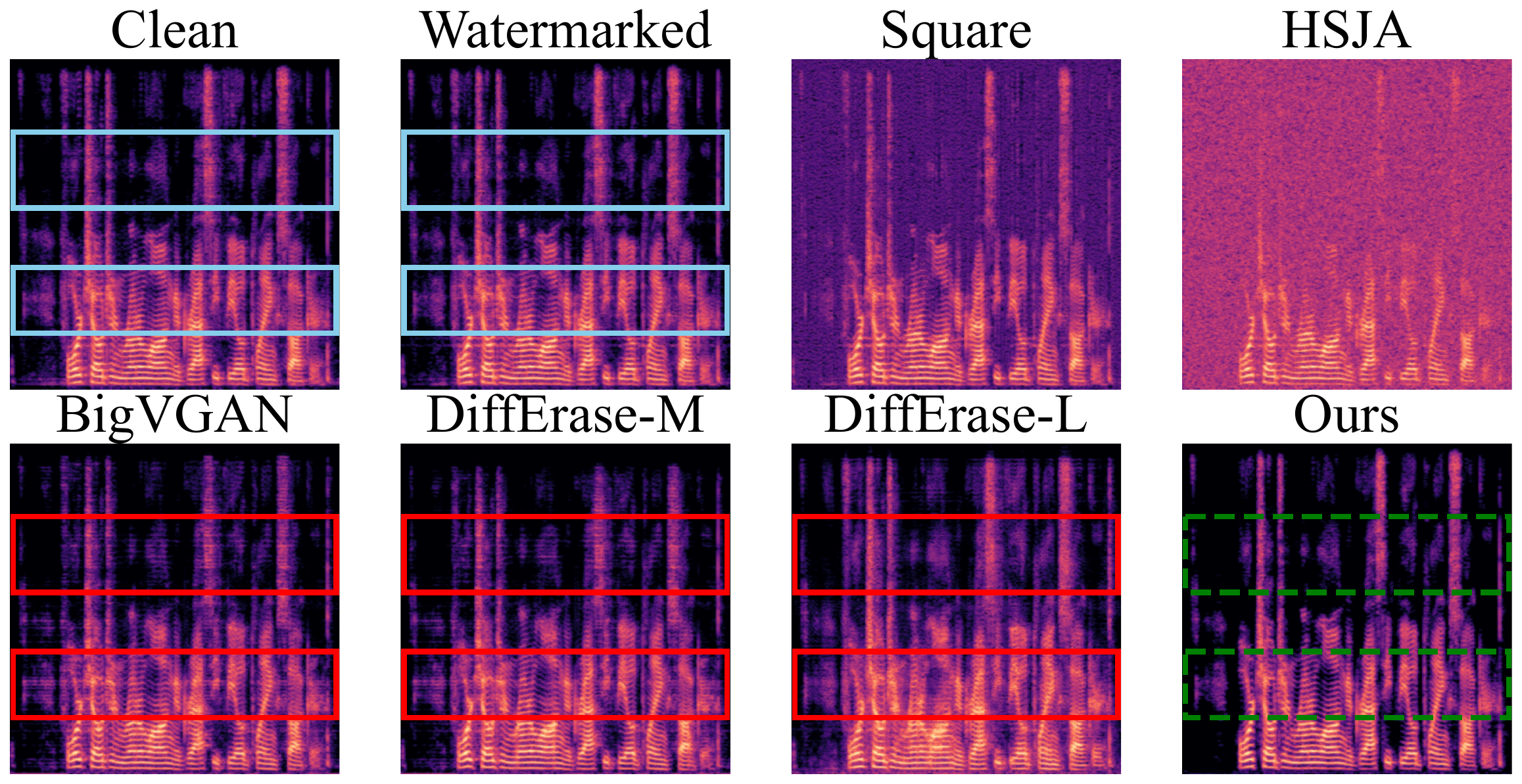}
    \caption{Attacked audio visualization (Timbre).}
    \label{fig:audio_visualization_timbre}
\end{figure*}

\begin{figure*}[t]
    \centering
    \includegraphics[width=\linewidth]{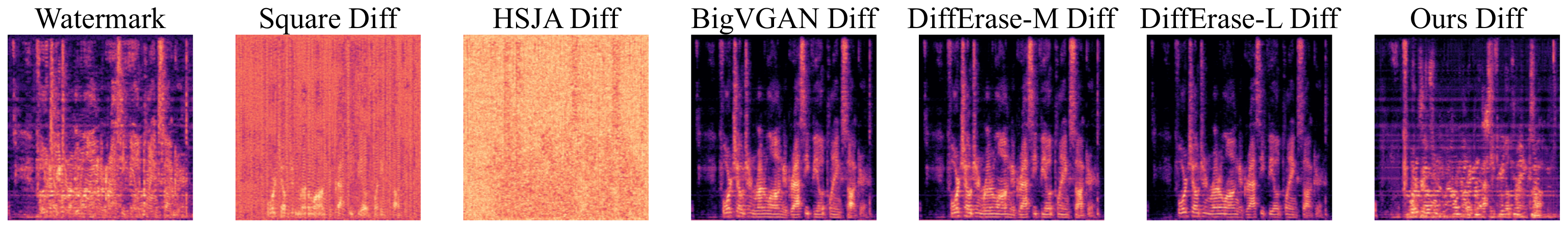}
    \caption{Pattern Visualization (Timbre).}
    \label{fig:audio_visualization_timbre_pattern_difference}
\end{figure*}

\begin{figure*}[t]
    \centering
    \includegraphics[width=0.75\linewidth]{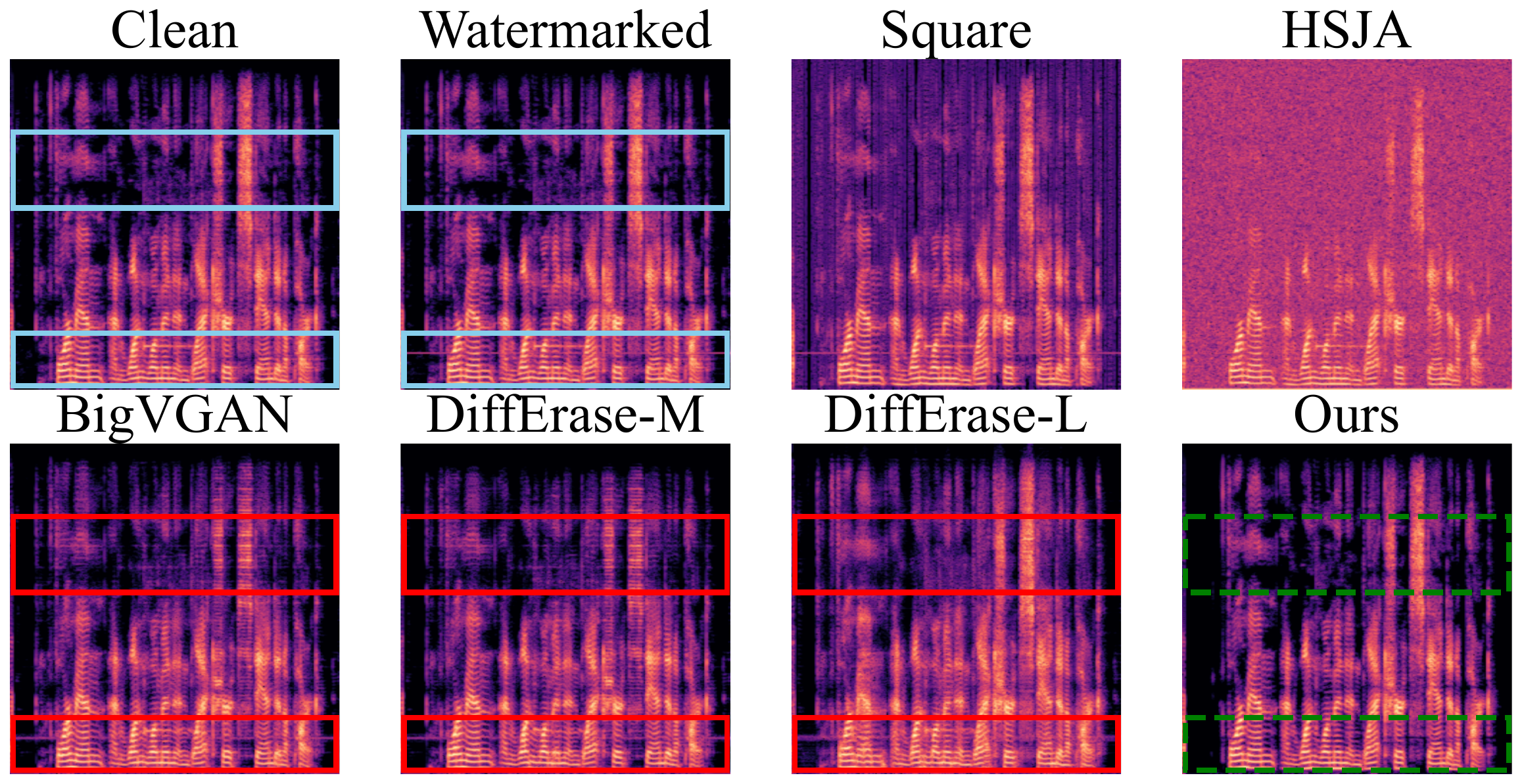}
    \caption{Attacked audio visualization (AudioMarkNet).}
    \label{fig:audio_visualization_audiomarknet}
\end{figure*}

\begin{figure*}[t]
    \centering
    \includegraphics[width=\linewidth]{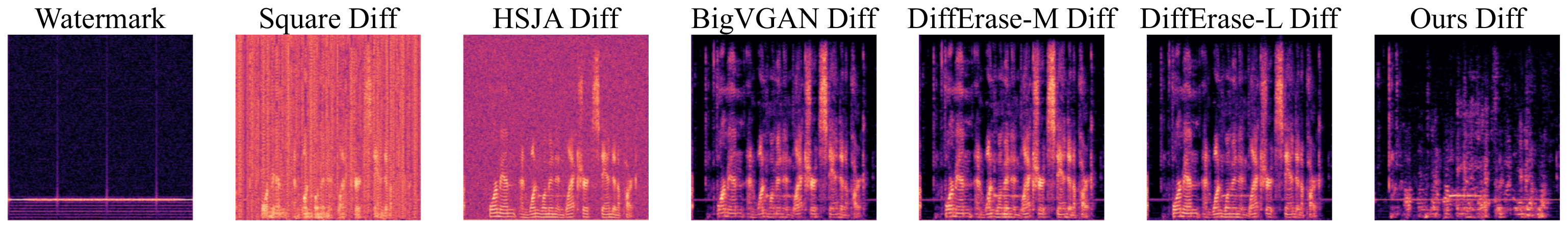}
    \caption{Pattern Visualization (AudioMarkNet).}
    \label{fig:audio_visualization_audiomarknet_pattern_difference}
\end{figure*}

Figures~\ref{fig:audio_visualization_audioseal_sup} and~\ref{fig:audio_visualization_audioseal_pattern_difference} present the audio and corresponding pattern spectrograms for AudioSeal. Similarly, Figures~\ref{fig:audio_visualization_wavmark} and~\ref{fig:audio_visualization_wavmark_pattern_difference}, Figures~\ref{fig:audio_visualization_timbre} and~\ref{fig:audio_visualization_timbre_pattern_difference}, and Figures~\ref{fig:audio_visualization_audiomarknet} and~\ref{fig:audio_visualization_audiomarknet_pattern_difference} show the results for WavMark, Timbre, and AudioMarkNet, respectively. The red solid boxes highlight regions with noticeable spectrogram distortions introduced by existing attacks, whereas the green dashed boxes indicate regions where our method better preserves the original speech structure.

We also report the watermark messages embedded in the representative audio samples. Because the watermark messages are randomly generated, we provide only the message corresponding to the representative sample selected for each watermarking method:

\begin{itemize}
\item AudioSeal watermark message: \\ \texttt{1010110110010000}.
\item WavMark watermark message: \\ \texttt{0110100011100001}.
\item Timbre watermark message: \\ \texttt{0110101110101000}.
\item AudioMarkNet watermark message:\\ \texttt{0000010011000111}.
\end{itemize}

\subsection{Additional Experiment on Ablation Study}
\label{sec:additional_experiment_on_ablation_study}

Table~\ref{tab:ablation_flickr} and Table~\ref{tab:ablation_wsj0} show the complete results for evaluating the effectiveness of adaptive scaling and the two artifact branches. The complement of the two patterns and adaptive scaling is necessary for suppressing diverse watermark artifacts.



\end{document}